\documentclass[useAMS,usenatbib]{mn2e}
\usepackage{psfig}
\usepackage{graphicx}

\def\cbeta{$c_{\beta}$}  

\def\arcsec{\hbox{$^{\prime\prime}$}}
\def\nii{[N {\sc ii}]}
\def\hi{H~{\sc i}}
\def\hii{H~{\sc ii}}
\def\sii{[S~{\sc ii}]}

\def\oii{[O~{\sc ii}]}

\def\heii{He~{\sc ii}}
\def\hei{He~{\sc i}}
\def\oiii{[O~{\sc iii}]}
\def\neiii{[Ne~{\sc iii}]}

\def\ha{H$\alpha$}
\def\hb{H$\beta$}
\def\hd{H$\delta$}   
\def\hg{H$\gamma$}   

\def\te{$T_e$}
\def\tenii{$T_e$[N~{\sc II}]}
\def\teoii{$T_e$[O~{\sc II}]}

\def\teoiii{$T_e$[O~{\sc III}]}

\def\ne{$N_e$}

\title[The chemical composition of the H~{\sc ii} regions in NGC 55]{NGC 55: a disc galaxy with flat abundance gradients\thanks{Based on observations obtained at the Gemini 
Observatory, which is operated by the Association of Universities for Research in Astronomy, 
Inc., under a cooperative agreement with the NSF on behalf of the Gemini partnership.}}

\author[Magrini, Gon\c calves \& Vajgel]{Laura Magrini$^{1}$\thanks{E-mail:laura@arcetri.astro.it}, Denise R. Gon\c calves$^{2}$, Bruna Vajgel$^{3}$ 
\\
  $^{1}$ INAF - Osservatorio Astrofisico di Arcetri, Largo E. Fermi 5, I-50125 Firenze, Italy\\  
  $^{2}$ Observat\'orio do Valongo, Universidade Federal do Rio de Janeiro, Ladeira Pedro Antonio 43, 20080-090 Rio de 
  Janeiro, Brazil\\
  $^{3}$ Observat\'orio Nacional - MCTI, Rua General Jos\'e Cristino 77, 20921-400 Rio de Janeiro, Brazil
}

\begin{document}

\date{Accepted ?. Received ?; in original form ?}

\pagerange{\pageref{firstpage}--\pageref{lastpage}} \pubyear{2013}

\maketitle

\label{firstpage}

\begin{abstract}
 
We present  new spectroscopic observations obtained with GMOS@Gemini-S of a sample of 25 \hii\ regions located in NGC~55, a late-type galaxy in the nearby Sculptor group. 
We derive physical conditions and chemical composition through the \te-method for 18 \hii\ regions, and strong-line abundances for 22 \hii\ regions. 
We provide abundances of He, O, N, Ne, S, Ar, finding a substantially homogenous composition in the ionised gas of the disc of NGC~55, with no trace of radial gradients. 
The oxygen abundances, both derived with \te- and strong-line methods, have similar mean values and similarly small dispersions: 12+$\log$(O/H)=8.13$\pm$0.18~dex with the former 
and  12+$\log$(O/H)=8.17$\pm$0.13~dex with the latter. 
The average metallicities and the flat gradients agree with previous studies of smaller samples of \hii\ regions and there is a qualitative agreement with the blue supergiant radial gradient as well. 
We investigate the origin of such flat gradients comparing NGC~55 with NGC~300, its companion galaxy, which is also twin of NGC~55 in terms of mass and luminosity. 
We suggest that the differences in the metal distributions in the two galaxies might be  related to the differences in their K-band surface density profile. The flatter profile of NGC~55 probably 
causes in this galaxy higher infall/outflow rates than in similar galaxies. This likely provokes a strong mixing of gas and a re-distribution of metals.  

\end{abstract}


\section{Introduction}

NGC~55 is the nearest edge-on galaxy at a distance of 2.34 Mpc \citep{k16} and it is 
 member of the nearby Sculptor group consisting of approximately 30 galaxies \citep{cote97, jerjen00} and being dominated by the spirals NGC~300 and NGC~253. 
The main properties of NGC~55 are listed in Table~\ref{tabNGC55}.

The nature of the NGC~55 galaxy has been debated for long time: its high inclination \citep[79$^{\circ}$;][]{puche91} has allowed different 
interpretation of its morphology. It has been sometimes defined as a late-type spiral galaxy \citep{ST87}, while in other works 
it has been considered as a dwarf irregular galaxy, similar to the Large Magellanic Cloud (LMC), as, e.g., \citet{dV61}. 
Following \citet{dV61} the main light concentration at visible wavelengths, which is offset from the geometric center of the galaxy, is a bar seen end-on. 
The structure of the disc of NGC~55 shows asymmetric extra-planar morphology \citep{ferguson96}. 
The extra-planar asymptotic giant branch (AGB) population is essentially old with ages of about 10~Gyr \citep{Davidge05}. 
\citet{tanaka11} studied  the structure and stellar populations of the Northern outer part of the stellar halo: 
from the stellar density maps they detected an asymmetrically disturbed, thick disc structure and possible remnants of merger events. 

Its interstellar medium (ISM) has been studied in several aspects: 
the neutral component \citep[e.g.,][]{hummel86, puche91, westmeier13}, the molecular component \citep[e.g.,][]{DH89,HD90},   and the 
ionised component \citep[e.g.,][]{WS83, hoopes96, ferguson96,tullmann03}. 
The star formation activity is located throughout the disc planar region of NGC~55, but there are also 
large quantities of gas off of the disc plane still forming stars. \citet{OD99} found shell structures and chimneys outside the planar regions that are signatures of supernovae
explosions and stellar winds. 
The composition of its exceptionally active population of \hii\ regions was studied in the past by 
\citet{WS83}, and more recently by \citet{tullmann03} and \citet{TR04} who studied
some regions, inside and outside the disc, respectively. 

The  radial metallicity gradient of NGC~55 was first outlined by the study of \hii\ regions of \citet{WS83} who found a substantially flat metallicity radial profile. A similar result  was obtained by \citet{Pilyugin14} in their re-analysis of the radial metallicity profiles of several late-type spiral galaxies including NGC~55. 
\citet{tikhonov05}  analysed the spatial distribution of the AGB and red giant branch (RGB) stars along the galactocentric radius of  NGC~55,  revealing again a very small metallicity gradient also for the older stellar populations. 
The absence of metallicity gradients might suggest a coherent formation of all the disc  or very efficient  mixing processes since
its formation.
From a sample of 12 B-type supergiant
stars, \citet{castro08} found a mean metallicity of $-$0.40~dex, a value quite close to the LMC metallicity
\citep[see, e.g.,][]{hunter07} and a flat gradient. 
The metallicities of the two extra planar \hii\ regions were found to be slightly lower than those of the central \hii\ regions, suggesting that 
they might have formed from material that did not
originate in the thin disc \citep{tullmann03}.

From a dynamical point of view, 
there are a number of works that confirm tidal interactions among the three pairs (NGC~55 and 300, 247 and 253, and 45 and 7793) of  major galaxies in Sculptor Group \citep[see, e.g.,][]{dV68, whiting99, westmeier13}. 
On top of the dynamical effects of the nearby galaxies on NGC~55, the kinematics of its central regions within the bar shows a gradient in radial velocities 
towards the galactic centre, which is due to flow of material along the bar \citep{CA88,westmeier13}. 
 
 \begin{table*}
\begin{minipage}{75mm}
	\caption{Properties of NGC~55}
	\label{tab:properties_NGC55}
	\begin{tabular}{lll} 
		\hline
		Parameter & Value & Reference \\
		\hline
		Type & SB(s)m & \citet{dV61} \\
		Centre &00h14m53s.6  -39$^{\circ}$11'47" & NED (J2000.0) \\
		Distance & 2.34$\pm$0.2 Mpc& \citet{k16} \\
		Total mass & 2.0$\pm$10$^{10}$M$_{\odot}$ &\citet{westmeier13} \\ 
		Inclination & 79$\pm$4$^{\circ}$ & \citet{puche91} \\
		Position angle & 109$^{\circ}$ & HyperLeda\\	
		\hline
	\end{tabular}
\label{tabNGC55}
\end{minipage}       
\end{table*}

The present work is part of our study of the structure and evolution of  Local Group and nearby galaxies through the spectroscopy 
of their emission-line populations \citep[see, e.g., ][]{magrini05,goncalves07, MG09, magrini09, goncalves12, goncalves14, stanghellini15}.
In this framework, we have carried on a deep spectroscopic campaign with the multi-object spectrograph GMOS@Gemini-South telescope of the strong-line emitters of NGC~55, 
as illustrated in Figure~\ref{fig_allsources}. The aim of the present work is to explore the distribution of abundances in this galaxy, studying \hii\  regions located in the disc --from the inner disc to the outskirts-- as well as extra planar regions. 

The paper is structured as follows: in Section~\ref{sec_obs} we describe the observations --imaging and spectroscopy-- and the data reduction process. 
In Section~\ref{sec_ana} we present  the spectroscopic analysis, whereas in Section~\ref{sec_grad} we describe the spatial distributions of the abundances. 
In Section~\ref{sec_dis} we discuss our results and compare them with those of its  companion galaxy NGC~300. In Section~\ref{sec_conclu} we give our conclusions.

\section[]{GMOS@Gemini-S: imaging and spectroscopy}
\label{sec_obs}

\begin{figure*} 
   \centering
   \includegraphics[width=16.5truecm]{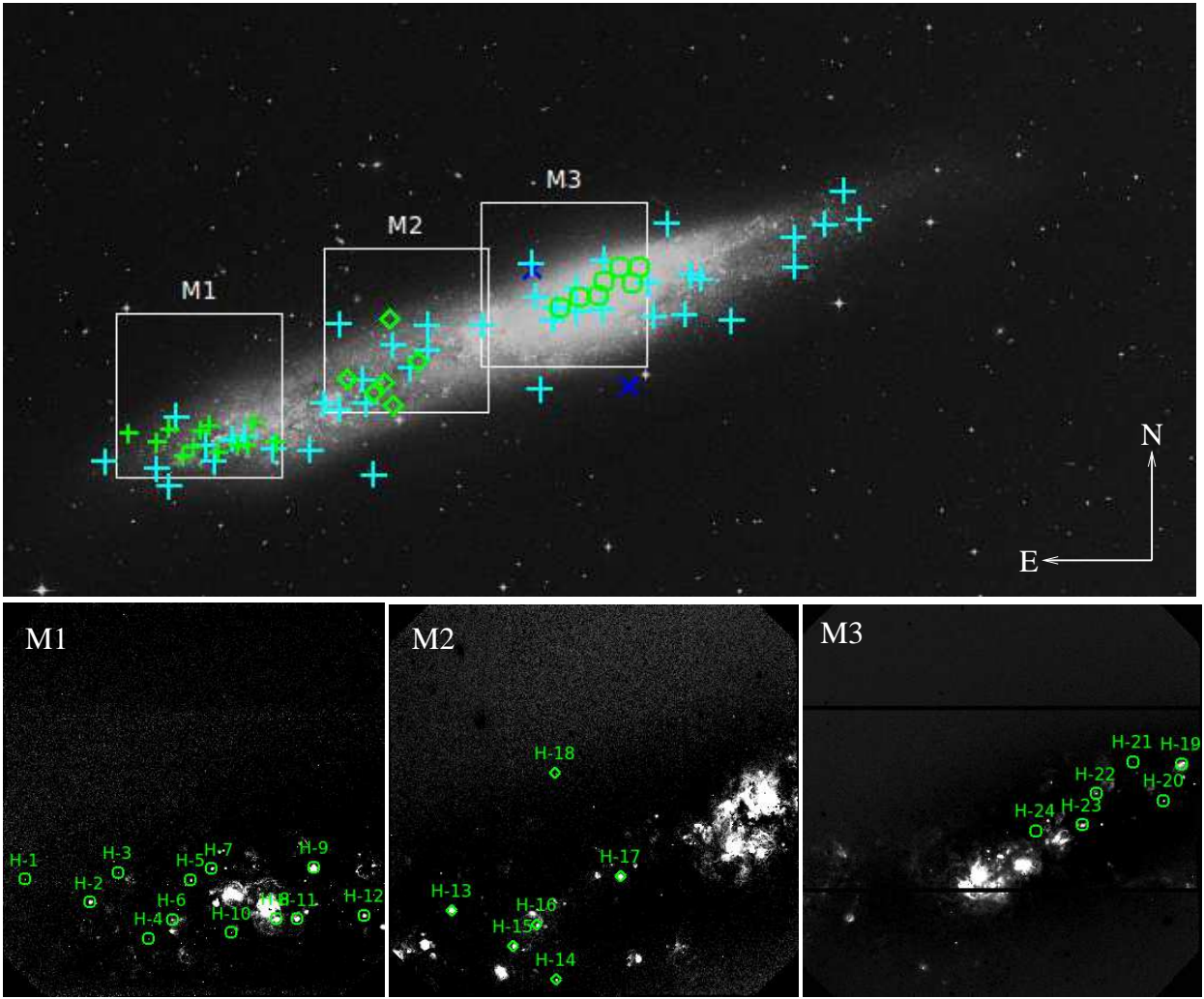} 
 \caption{ 
 {\it Top:} 
 HSDSS IIIaJ4680\AA\ image of NGC~55. The entire image is 30$\times$15~arcmin$^2$, and it is centred at RA=00:15:12.27 and DEC=-39:12:57.89. Superposed to it the three GMOS-S FoV we observed  (masks M1, M2 and M3), of 5.5$\times$5.5~arcmin$^2$ each, are highlighted. 
 The green symbols (plus, diamond and circle) are our \hii\ regions located within these three FoV, respectively. Other symbols are as follows: cyan plus, the X-rays sources up to D$_{25}$ from \citet{stobbart06}; and the two blue crosses indicate the extra-planar \hii\ regions studied by \citet{tullmann03}.
 {\it Bottom:} 
Our GMOS-S continuum subtracted images (H$\alpha$-H$\alpha$C) of the three FoV. The \hii\ regions we discuss in this paper are identified, following the IDs of 
Table~\ref{tab_objid}. The orientation, North to the top and East to the left is the same in all the panels. }
   \label{fig_allsources}
\end{figure*}

\begin{table}
\centering
\begin{minipage}{75mm}
{\scriptsize  
\caption{GMOS-S mask identification, classification and coordinates of the \ha\ line-emitters selected from the GMOS-S pre-imaging. M1, M2 and M3 stands for the masks ID. The object classification shown is based on the follow-up spectroscopic analysis of the present study.}
\begin{tabular}{@{}llllll@{}}
\hline
MaskId & Field-ID & Class   & RA           & Dec  	   \\
         &     &     & J2000.0      & J2000.0	   \\        
\hline    
M1S1 & H-1 & HIIr       &  00:16:14.93 & -39:15:51.62    \\   	
M1S4 &H-2 & HIIr       &  00:16:10.12 & -39:16:11.49    \\	
M1S5 &NGC~55 StSy-1  & SySt    &  00:16:07.25 & -39:16:31.91 \\	
M1S6 &H-3 & HIIr       &  00:16:07.99 & -39:15:45.94    \\	
M1S7 &H-4 & HIIr       &  00:16:05.73 & -39:16:42.96    \\	
M1S8 &H-5 & HIIr       &  00:16:02.59 & -39:15:52.35    \\	
M1S10 &H-6 & HIIr       &  00:16:03.96 & -39:16:26.61    \\	
M1S12 &H-7 & HIIr       &  00:16:01.04 & -39:15:42.52    \\	
M1S13 &H-8 & HIIr       &  00:15:56.26 & -39:16:25.79    \\	
M1S15 &H-9 & HIIr       &  00:15:53.42 & -39:15:41.12   \\	
M1S16 &H-10 & HIIr       &  00:15:59.58 & -39:16:37.63    \\	
M1S17 &H-11 & HIIr       &  00:15:54.64 & -39:16:25.90    \\	
M1S18 &H-12 & HIIr       &  00:15:49.66 & -39:16:23.08    \\	
\\    
M2S1 & H-13 & HIIr       &  00:15:36.94 & -39:14:21.08     \\		
M2S2 & NGC~55 SySt-2 & SySt       &  00:15:39.02 & -39:14:40.60  \\     
M2S3 &  H-14& HIIr       &  00:15:29.36 & -39:15:18.69      \\		
M2S5 & H-15 & HIIr       &  00:15:32.49 & -39:14:50.50     \\		
M2S6 &  H-16& HIIr       &  00:15:30.75 & -39:14:32.54     \\		
M2S14 & H-17 & HIIr       &  00:15:24.71 & -39:13:51.88     \\		
M2S15 & H-18 & HIIr       &  00:15:29.41 & -39:12:25.09   \\	        
\\
M3S1  & H-19& HIIr  	    &  00:14:46.02 & -39:11:01.79     \\ 
M3S2  & H-20& HIIr  	    &  00:14:47.32 & -39:11:32.85     \\ 
M3S3  & H-21& HIIr  	    &  00:14:49.52 & -39:10:59.99     \\ 
M3S4  & H-22& HIIr  	    &  00:14:52.18 & -39:11:26.70     \\ 
M3S5  & H-23& HIIr  	    &  00:14:53.18 & -39:11:53.30     \\ 
M3S7  & H-24& HIIr  	    &  00:14:56.52 & -39:11:58.05     \\ 
M3S8  & NGC~55 StSy-3 & SySt &  00:14:58.61 & -39:11:59.14    \\ 
M3S10  & H-25& HIIr 	&  00:14:59.94 & -39:12:14.97     \\	 
\hline					  
\end{tabular}
}
\end{minipage}
\label{tab_objid}
\end{table}

The data analysed in the present paper were obtained with the Gemini Multi-Object Spectrographs (GMOS) at Gemini South telescope in 2012 and 2013. The two programs through which the data were taken are GS-2012B-Q-10 and GS-2013B-Q-12, with D. R. Gon\c calves as Principal Investigator (PI) in both cases. In total we observed three fields of view of GMOS-S, each of 5.5\arcmin$\times$5.5\arcmin. In the following, we refer to the three fields as M1, M2 and M3 (see Figure~ \ref{fig_allsources}).

\subsection*{Pre-Imaging}

We obtained the pre-imaging of NGC~55 with the GMOS-S camera in queue mode on the 28$^{th}$ (M1) and 27$^{th}$ (M2, M3) of August 2012. 
We used the on- and off-band imaging technique to identify the strongest H$\alpha$ line emitters. Their location is shown in Figure~\ref{fig_allsources}. 
For the three fields of view  (FoV) the on-band H$\alpha$ images were sub-divided in 3 exposures of 60~s each, while the three off-band (the continuum of H$\alpha$) H$\alpha$C sub-exposures were of 120~s each. 
These narrow-band filters have central $\lambda$ ($\lambda$ interval) of  656nm (654-661nm) and 662nm (659-665nm) for H$\alpha$ and  H$\alpha$C, respectively. 
The location of the three fields (5.5\arcmin$\times$5.5~\arcmin) is shown in Figure~\ref{fig_allsources}, and the central coordinates of each field are: for M1  R.A. 00:16:02.62 and Dec. -39:14:41.77;
for M2 R.A. 00:15:26.59 and Dec. -39:12:49.08; and for M3 R.A. 00:14:59.06 and Dec. -39:11:30.97. 
During the pre-imaging observations, the seeing varied from 0.9'' to 1\farcs0.

In Table~\ref{tab_objid} we give the coordinates of the observed emission-line objects (25 \hii\ regions, HIIr, and 3 candidate symbiotic systems, SySt). We based the classification of the nebulae on the analysis of their spectra, which will be introduced in the following sections. We discriminate SySts from \hii\ regions on the bases of the presence of absorption features and continuum of late-type M giants, of the 
strong nebular emission lines of Balmer \hi, \heii, the simultaneous presence of forbidden lines of low- and high-ionization, like \oii, \neiii, \oiii\,  \citep{bel00} and of the Raman scattered line at $\lambda$6825\AA, a signature almost exclusively seen in symbiotic stars \citep{schmid89, bel00}.

In Figure~\ref{fig_allsources} the positions of our \hii\ regions are shown in contrast with objects investigated in previous works:  
X-rays sources up to D$_{25}$\footnote{D$_{25}$ is the diameter that corresponds to a surface brightness of 25 mag/arcsec$^2$} from \citet{stobbart06}; and the extra-planar \hii\ regions from \citet{tullmann03}. 

\subsection*{Spectroscopy}

The spectroscopic observations were obtained in queue mode with two gratings, R400+G5305 (red) and B600 (blue). For the mask M1, 
the spectra were taken on the 22$^{nd}$ (blue) and 23$^{th}$ (red) of December 2012. The spectroscopy of M2 and M3, on the other hand, were obtained about one year later. The red  spectra of M2 on the 29$^{th}$ of September, while the blue counterpart was observed on the 13$^{th}$ of October, in 2013. As for M3, 
the red and blue spectra were taken on the nights of 27$^{th}$ and 29$^{th}$ of December of 2013.

We obtained three exposures per mask and grating --with 3$\times$1,430s for  M1 and 3$\times$1,390s for M2 and M3 
both in blue and red.  For technical reasons, only one exposure of the blue spectra of M3 and two of the red spectra of M2 were useful for science. 
In all the cases, we combined the good quality exposures to increase the signal-to-noise ratio (SNR) of the spectra and to remove cosmic rays. 

\begin{figure} 
   \centering
   \includegraphics[width=8.5truecm]{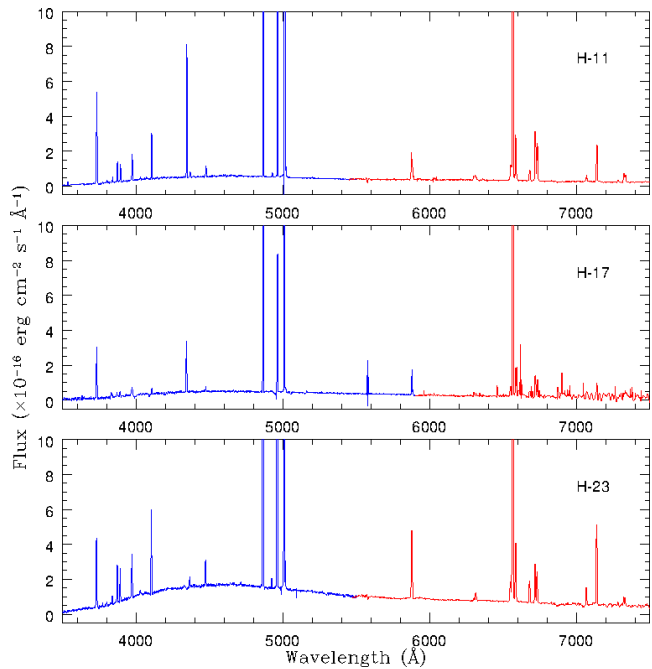}
 \caption{Sample of our GMOS spectra, one per FoV, M1, M2 and M3, for the \hii\ regions H-11, H-17 and H-23, respectively. The CCD gaps in the blue part of H-23's spectrum were masked, since in this case only one spectrum was good enough for science, so the CCD gaps could not be erased by the combination of different frames. Also note the red spectrum of H-17, noisier than in of the other two \hii\ regions due to the fact that 2, instead of 3 spectra were combined to obtain this one.}
   \label{fig_spectra}
\end{figure}

For most of the spectra the effective blue plus red spectral coverage range from 3500~\AA\ to 9500~\AA, in several cases allowing a significant overlap of the spectra. Only few lines were measured with wavelength longer than ~7200~\AA, because of the poor (wavelength + flux) calibration at the red end of the spectra  (see Figure~\ref{fig_spectra}). We avoided the possibility of important emission-lines to fall in the gap between the three CCDs, by slightly varying the central wavelength of the disperser from one exposure to another. To do this, we centred the red grating R400+G5305 at 750 $\pm 10$~nm and the blue one B600 at 460 $\pm 10$~nm. 

The masks were built with slit widths of 1\arcsec\ and with varying lengths to include portions of sky in each slit for a proper local sky-subtraction.  
The spectroscopic observations were spatially$\times$spectrally binned. The final spatial scale and reciprocal dispersions of the spectra were: 0\farcs144 and 0.09~nm per pixel, in blue; and 0\farcs144 and 0.134~nm per pixel, in red. 

Following the usual procedure with GMOS for wavelength calibration, we obtained CuAr lamp exposures with both grating configurations, either the day before and after the science exposures.  
The spectrophotometric standard LTT7379 \citep{hamuy92} was observed with the same 
instrumental setups as for science exposures, on September 2$^{nd}$ 2013, and used for  flux calibration of the three masks, since no standards were obtained near the observation of M1, in 2012. In Figure~\ref{fig_spectra} we show a sample with fully reduced and calibrated GMOS spectra, one spectrum per field, on which it is straightforward to see the 
quality of our data. 

Data were reduced and calibrated in the standard way by using the Gemini {\sc gmos data
  reduction script} and {\sc long-slit} tasks, both being part of
{\sc IRAF}\footnote{IRAF is distributed by the National Optical Astronomy
  Observatory, which is operated by the Association of Universities
  for Research in Astronomy (AURA) under cooperative agreement with
  the National Science Foundation.}

\section[]{Spectral analysis and determination of the physical and chemical properties}
\label{sec_ana}

We measured the emission-line fluxes and their errors with the package {\sc SPLOT}
of {\sc IRAF}. Errors take into account
the statistical errors in the measurement of the fluxes and the 
systematic errors (flux calibrations, background determination
and sky subtraction). 

We corrected the observed line fluxes for the effect of 
the interstellar extinction using the extinction law of \citet{mathis90} with
R$_{\rm V}$=3.1  (which is assumed to be  constant as in our Galaxy, but might slightly vary from galaxy to galaxy, see, e.g.,  \citet{clayton15})  and the individual reddening of each \hii\ region given by  the  
\cbeta, i.e., the logarithmic difference between the observed and theoretical \hb\ fluxes. 
Since \hd\ and \hg\ lines are fainter than the  \ha\ and \hb\ ones, and located in the bluer part of the spectra, they are  consequently  affected by larger uncertainties.  
We thus determined  \cbeta\ comparing the 
observed Balmer I(\ha)/I(\hb) ratio with its theoretical
value, 2.87 \citep[case B. c.f., ][]{osterbrock06}. 
We used lines in common between the blue and red part of the spectra to put the two spectral ranges on the same absolute flux scale. 
The average scaling factor applied to the fluxes of  the red spectra is 0.78. 

In   Table~B1 of the Appendix we present the measured and extinction corrected intensities --both normalised to H$\beta$=100-- 
of the emission lines of the 25 \hii\ regions listed in Table~\ref{tab_objid}. The spectra of the candidate symbiotic systems will be discussed in a forthcoming paper.    

The method to derive the physical and chemical condition in  \hii\ regions is described 
in the previous papers of this series, as for instance \citet{MG09, goncalves12, goncalves14}: 
the {\sc temden} and {\sc ionic} tasks of {\sc IRAF}\footnote{The atomic data source is the that of 
{\sc analysis/nebular -- IRAF};  
http://stsdas.stsci.edu/cgi-bin/gethelp.cgi?at\_data.hlp} 
are used to derive first electron temperature (from the ratio of the \oiii\ lines at 500.7~nm and 436.3~nm),  and density
(from the lines of the \sii\ doublet at 671.7~nm and 673.1~nm) of the gas, then ionic abundances.
Ionic abundances are combined with the ionisation correction 
factors for \hii\ regions from \citet{izotov06} to obtain the total abundances. 

Often, literature studies represent  \hii\ regions with a two-zone ionisation structure characterised by two different electron temperature, for the \oii\ and \oiii\ emitting regions. 
If available \teoii\ --or equivalently  \tenii\ -- are used for the \oii\ zone, while \teoiii\ is adopted in the \oiii\ emitting zone. When  \teoii\ or     \tenii\ are not measured, the relation, based on 
the photoionisation models of \citet{stasinska82},   
\teoii=0.7$\times$\teoiii+3000 K (or a similar one) is adopted. If we consider valid without errors the linear relation between \teoii\ and \teoiii, 
in the temperature range of our \hii\ regions, the use of a single temperature (the measured one, i.e., \teoiii) might imply a maximum  underestimation 
of about $\sim$700 K and up to a maximum overestimation of $\sim$400 K of the temperature adopted for the \oii\ ionic abundance. We have tested the effect of adopting two different temperatures 
and, for most regions, it has negligible effect on the total oxygen abundance --of the order 0.01-0.03 dex-- since \oiii\ ionic abundance is on 
average a factor 10 larger than \oii\ ionic abundance, and consequently it is the ionic fraction that  contributes the most to the total O/H. 
Thus we use only the measured  \te\oiii\  for the calculation of the abundances of both low- and high-ionisation species.

The abundances of \hei\ and \heii\ were computed using the equations
of \citet{benjamin99} in two density regimes, that
is \ne\ $>$ 1,000 and $\le$ 1,000 cm$^{-3}$. Clegg's collisional populations
were taken into account \citep{clegg87}.

 The results of physical and chemical properties of the \hii\ regions are shown in Table~\ref{tabu2}. 
The \te\ has been measured in 18 regions and ranges from 8,600 to 12,000~K; \ne\ is available in 10 regions, while an upper limit is measured in other 6 regions. 
The measurements of electron densities range from 100 to 600 cm$^{-3}$, typical values of \hii\ regions, with a higher density of 1400 cm$^{-3}$ in H-13. 
Errors on final abundances take into account 
the errors on \te\, on \ne\ and on the line fluxes. 
The average oxygen abundance determined with the \te\ method, excluding lower limit abundances,  is 12+$\log$(O/H)=8.13$\pm$0.18. For the other elements we obtain average values: 12+$\log$(N/H)=7.18$\pm$0.28, 12+$\log$(Ne/H)=6.81$\pm$0.14, 12+$\log$(Ar/H)=6.00$\pm$0.25,
and 12+$\log$(S/H)=6.04$\pm$0.25.  
Helium abundance is quite uniform too. Its mean value, linearly expressed, is  He/H=0.092$\pm$0.019.
The mean O/H is in good agreement with the metallicity measured by \citet{tullmann03} in their \hii\ region located in the disc-halo transition of NGC~55 (8.05$\pm$0.10 with the \te-method). The two extra-planar \hii\ regions of \citet{tullmann03} are instead slightly metal poorer (7.77 and 7.81) than our average O/H, but they are still consistent with the composition of some individual \hii\ regions, as for instance H-19, H-20, H-21.  
This might be an indication of incomplete mixing in the disc of NGC~55. 

\begin{table*}
\begin{minipage}{1000mm}
{\small  
\caption{Electron temperatures, electron densities, ionic and total abundance of the \hii\ regions.}
\begin{tabular}{@{}lrrrrrrrrr@{}}
\hline
Diagnostic                     & H-1		         & H-2	               &   H-5		  & H-6	      		       & H-7		 & H-11  	          &  H-12	       & H-13    	 	  & H-14  \\
\hline
$T_{e}$[O~{\sc iii}](K)        &  12000$\pm$200   & 9200$\pm$150    &  9300$\pm$150     & 11500$\pm$200      &12000$\pm$200  	& 8700$\pm$150&11000$\pm$200   &10700$\pm$200  &10800$\pm$200     \\
$N_{e}$~[S~{\sc ii}](cm$^{-3}$)&  -    	         &  $<100$ 	       & 100$\pm$50         &  $<100$  	                &200$\pm$50        	 & $<$100 & 600$\pm$100       & 1400$\pm$200    &250$\pm$50   \\
He~{\sc i}/H 	               & 0.14  	  		& 0.07	    	       &  0.09                     & 0.07                         &  0.11                    	  & 0.10  	          & 0.075	       		& 0.07	      & 0.10     \\
He/H  	                        &0.14$\pm$0.03        &0.07$\pm$0.01      &  0.09$\pm$0.01    &0.07$\pm$0.01          &0.11$\pm$0.05    	&0.10$\pm$0.05  	& 0.075$\pm$0.05  &0.07$\pm$0.02   &  0.10$\pm$0.05 \\
O~{\sc ii}/H 	               & 2.1e-06	  	        & 3.2e-05	    	       &  1.9e-05               & 5.4e-05         		&  1.4e-05 	   	  & 6.1e-05	 	  & 2.2e-05	       & -              & 1.6e-05	       	   \\
O~{\sc iii}/H 	               & 1.3e-04	                & 1.1e-04	               &  1.5e-04                & 1.9e-05         		&  7.9e-05 	   	  & 2.0e-04	 	  & 1.1e-04	       & 8.9e-05      & 9.8e-05	        	  \\
ICF(O)             	       	       & 1.2  	                & 1.0	                       &  1.0                       & 1.0                		&  1.0                  	  & 1.0  	          & 1.0	       	       & -            	   & 1.0  	     		  \\
O/H  	           	       	       & 1.6e-04	                & 1.4e-04	              &  1.7e-04                & 3.5e-05         		&  9.3e-05 	   	  & 2.6e-04	          & 1.3e-04	       & $>$8.9e-05    & 1.1e-04	          \\
12+log(O/H)      	       & 8.21$\pm$0.07       &8.15$\pm$0.06     &  8.23$\pm$0.06    & 7.90$\pm$0.10    	&  7.97$\pm$0.12       & 8.42$\pm$0.12& 8.12$\pm$0.08  & $>$7.95         & 8.06$\pm$0.07   	   		\\
N~{\sc ii}/H  	               & 8.9e-07	                & 3.5e-06	              &  2.8e-06                & 3.9e-06         		&  1.5e-06 	  	    & 2.9e-06	          & 1.4e-06	       & 1.17e-06           & 2.2e-06		   		 \\
ICF(N)             	       	       & 45.0 	                & 4.5	                       &  7.6                       & 1.6             		&  6.2            	     & 4.4  	          & 5.5	               & -	     	& 6.5  	       	\\
N/H    	           	       & 3.9e-05	                & 1.5e-05	              &  2.1e-05                 & 6.2e-06         		&  9.2e-06 	     & 1.2e-05	          & 7.6e-06	       &  -        & 7.7e-05               		\\
12+log(N/H)      	       & 7.58$\pm$0.15       & 7.18$\pm$0.15    &  7.32$\pm$0.15     & 6.80$\pm$0.15     	&  6.96$\pm$0.12    & 7.10$\pm$0.15& 6.88$\pm$0.12   & -             & 7.16$\pm$0.12     		  \\
Ne~{\sc iii}/H                    & 5.1e-06	                & 4.7e-06	     	     &  5.5e-06                 & -         			   &  1.0e-05 	       & 8.0e-06	          & 5.2e-06	       & -         	& 8.1e-06	        	\\
ICF(Ne)            	               & 1.0  	                & 1.2	    		     &  1.0                        & -             			&  1.1            	     & 1.2  	          & 1.1	       	        & -	         	& 1.0 	        	\\
Ne/H   	           	       & 5.1e-06	                & 5.5e-06	     	     &  5.5e-06                 & -         			&  1.1e-05 	      & 9.5e-06	          & 5.7e-06	       & -            & 8.5e-06	      		 \\
12+log(Ne/H)     	       & 6.70$\pm$0.14      & 6.74$\pm$0.15    &  6.75$\pm$0.15     & -     				&  7.03$\pm$0.04    & 6.7$\pm$0.14 & 6.76$\pm$0.14      & -               & 6.93$\pm$0.15        		 \\
Ar~{\sc iii}/H                     & -	                        & 9.4e-07	             &  8.8e-07                 & 2.9e-07         		&  5.4e-07 	  	 & 1.4e-06	          & 7.5e-07	       & -               	   	& -	             \\
ICF(Ar)            	               & -  	                	       & 1.1	                     &  1.2                        & 1.2             		&  1.1            	     & 1.1  	          & 1.1	       		& -	         	   	& -	         \\
Ar/H 	           	               & -	                        & 1.0e-06	            &  1.0e-06                  & 3.5e-07         		&  6.0e-07 	 	  & 1.5e-06	          & 8.2e-07	       & -          	& -	                     \\
12+log(Ar/H)     	       & -                             & 6.00$\pm$0.20   &  6.00$\pm$0.20      & 5.53$\pm$0.20     	&  5.78$\pm$0.22        & 6.20$\pm$0.25& 5.90$\pm$0.23  & -               	& -	           \\
S~{\sc ii}/H 	               & -	                        & 1.2e-06	            &  8.5e-07                  & 7.2e-07         		&  3.4e-07 	 	   & 8.7e-07	           & 4.2e-07	       & 2.4e-07    	& 6.3e-07	          \\
S~{\sc iii}/H 	               & -	                        & -	                     &  -                             & -           			&  5.1e-07          	  &-				&-				&-   	& -	          \\
ICF(S) 	           	       & - 	                        & 1.3	                     &  2.0                         & 1.0             		&  1.7            	          & 1.3  	           	& 1.5	       			& -	        	& 1.7    \\
S/H 	           	               & -	                        & 1.6e-06	            &  1.7e-06                  & 7.2e-07         		&  1.4e-06 	 	  & 1.2e-06	           & 6.5e-07	       & -            & 1.1e-06	       		    \\
12+log(S/H)      	       & -                             & 6.19$\pm$0.30   &  6.22$\pm$0.30      & 5.90$\pm$0.30     	&  6.16$\pm$0.28         & 6.10$\pm$0.30 & 5.80$\pm$0.30  & -       & 6.05$\pm$0.30         \\
\hline
\end{tabular}
}
\end{minipage}
\label{tabu1}
\end{table*}

\begin{table*}
\begin{minipage}{1000mm}
{\small  
\contcaption{}
\begin{tabular}{@{}lrrrrrrrrr@{}}
\hline
Diagnostic                      	          	   & H-15		  & H-17	  			& H-18  	 		& H-19	   	& H-20			&  H-21   			& H-22 	   		& H-23	    		& H-25 \\
\hline
$T_{e}$[O~{\sc iii}](K)            		&8600$\pm$150	  &10000$\pm$150  &12000$\pm$500 &12000$\pm$200    &12400$\pm$300	&10500$\pm$200	&9900$\pm$150	   &9100$\pm$150    &9600$\pm$150 \\
$N_{e}$~[S~{\sc ii}](cm$^{-3}$)       	& 100$\pm$50     & $<$100  	        & -		  	   & 300$\pm$100        &150$\pm$50  & $<$100	 	        & $<$100		           &150$\pm$50        &200$\pm$50  \\
He~{\sc i}/H 	                   		& 0.075 	           &  0.07	                &  0.10 	            & 0.08	                   &   0.10	& 0.09   	                & 0.095 	                    & 0.094	                & 0.115	    	 \\
He/H  	                          		& 0.075$\pm$0.02     & 0.07$\pm$0.02	   &  0.10$\pm$0.05 & 0.075$\pm$0.02     & 0.07$\pm$0.02   &0.09$\pm$0.01      &0.095$\pm$0.01         &0.094$\pm$0.01  & 0.115$\pm$0.03    \\
O~{\sc ii}/H 	                	        		& 2.6e-05	           &  4.2e-05               &  -	  		    & 2.0e-06	           & 1.3e-05	& 1.5e-05			& 1.0e-05	   		   & 1.6e-05      		& 4.2e-06  \\
O~{\sc iii}/H 	                          	& 2.1e-04	           &  1.1e-04               &  1.7e-04	    & 9.1e-05	           & 6.0e-05	 & 4.2e-05			& 1.3e-04	   		   & 1.6e-04      		& 2.4e-04         \\
ICF(O)             	                        		& 1.0 	           &  1.0	                 &  - 	                     & 1.0	                   & 1.0		  & 1.0   			& 1.0 	   		   & 1.0	    		& 1.0	       \\
O/H  	           	                     		& 2.3e-04	            &  1.6e-04              &  $>$1.7e-04	    & 9.3e-05	           & 7.3e-05	 & 5.7e-05			& 1.4e-04	   		   & 1.8e-04      		& 2.4e-04     \\
12+log(O/H)      	      			& 8.36$\pm$0.12 &  8.20$\pm$0.07  &  $>$8.25           & 7.70e$\pm$0.12  & 7.86$\pm$0.13    & 7.80$\pm$0.12 	&8.16$\pm$0.08    	   & 8.25$\pm$0.08 	& 8.38$\pm$0.12  \\
N~{\sc ii}/H  	           			& -	     			&  3.8e-06            &  -	                 &3.9e-07	    		&1.4e-06		& 3.0e-06	   		& 2.6e-06    		& 1.9e-06            & 1.9e-06 \\
ICF(N)             	              			& -	     			&  4.0	            &  - 	                 &34.2  	                 & 5.4		  & 4.0   			& 11.4	   		& 9.1	    			& 41.4	 \\
N/H    	           	        			& -                               & 1.4e-05	     & -	     		&  1.5e-05            	&7.2e-06	          & 1.2e-05	       	& 2.9e-05	   		& 1.6e-05      	&	 7.7e-05 \\
12+log(N/H)      	       			&-				 & 7.17$\pm$0.20 &-     			&  7.10$\pm$0.13  	& 6.86$\pm$0.10  & 7.10$\pm$0.12  	& 7.46$\pm$0.15   	& 7.22$\pm$0.10 	& 7.88$\pm$0.20    \\
Ne~{\sc iii}/H                      		& 6.9e-06	            	&  7.9e-06              &  -	                      & 4.8e-06	            & 3.8e-06	& -				& 5.0e-06	   		& 8.6e-06      		& 7.8e-06  \\
ICF(Ne)            	                          	&  1.2                     	&  1.2	               &-			  & 1.0	                    & 1.1		& -   				& 1.0 	   		& 1.0	    			& 1.0	    \\
Ne/H   	           	                  	& 7.1e-06	           	&  9.8e-06               &  -	                      &  4.8e-06	           & 4.2e-06	& -				& 5.0e-06	   		& 8.6e-06      		& 7.8e-06  \\
12+log(Ne/H)     	       			& 6.85$\pm$0.14    	&  7.00$\pm$0.14   &  -                        & 6.66$\pm$0.15    & 6.62$\pm$0.14 & -				& 6.70$\pm$0.15   	& 6.93$\pm$0.15 	& 6.89$\pm$0.15      \\
Ar~{\sc iii}/H                       		& -	                   	&  7.2e-07               &  -	                      & 4.6e-07	           & 6.0e-07	& 8.7e-07			& -	   			& 1.4e-06		      	& 1.2e-06     \\
ICF(Ar)            	                 	        &-  	                 	&  1.1 	                      & -			   &2.8	                   & 1.1		 & 1.1   			& - 	  	 		& 1.3		    		& 3.2	     \\
Ar/H 	           	               	                 & -	                   	&  7.6e-07               &  -	                      & 1.3e-06	           & 6.6e-07	& 9.2e-07			& -	   			& 1.8e-06		      	& 3.9e-06     \\
12+log(Ar/H)     	                         & -                           	&  5.88$\pm$0.24   &  -                         & 6.10$\pm$0.17    & 5.82$\pm$0.23   & 6.00$\pm$0.24 	&   		& 6.26$\pm$0.17 	& 6.59$\pm$0.23 \\
S~{\sc ii}/H 	                         		& 7.3e-07	          	&  5.3e-07               &  -	                      & 9.8e-08	           & 3.0e-07	& 6.2e-07			& 4.6e-07	   		& 3.0e-07     	 	& 6.1e-07   \\
ICF(S) 	           	       	                 & 2.0 	                  &  1.2	                &                             & 7.8	                   & 1.5		& 1.2   			& 2.8 	   		& 2.3	    			& 9.4	     \\
S/H 	           	                        		& 1.5	e-06                  	&  6.5e-07               &  -	                      & 7.7e-07	          & 4.5e-07	 & 7.8e-07			& 1.3e-06		   	& 6.9e-07      		& 5.8e-06 \\
12+log(S/H)      	      		     	& 6.20$\pm$0.30    	&  5.80$\pm$0.30   &  -                          & 5.90$\pm$0.30   & 5.65$\pm$0.30  & 5.90$\pm$0.30 	& 6.11$\pm$0.27      & 5.84$\pm$0.27 	& 6.76$\pm$0.30   \\
\hline
\end{tabular}
}
\end{minipage}       
\label{tabu2}
\end{table*}

\subsection{Strong-line metallicities} 

We computed oxygen abundances using strong-line methods to increase the number of regions with a determined metallicity. 
These methods are based on the intensities of lines that are usually easy to measure because  they are much stronger than the lines used as \te\ diagnostic \citep[c.f.][for a complete discussion and comparison among the methods]{ac15}.  
The strong-line ratios can be calibrated in two different ways: using photoionisation models or using abundances of \hii\ regions obtained through the \te-method.  Since the empirical calibration works better in the low metallicity regime, 
we have used the new calibrations based on \te-method abundances by \citet[][hereafter M13]{marino13} of the two well-known indices, N2=$\log$(\nii/\ha) and O3N2=$\log$[(\oiii/\hb)(\nii/\ha)]. The results are shown in Table~\ref{tab_strong} where we present the galactocentric distances, O/H from the \te-method, the metallicities derived with the M31's N2 and O3N2 indices, and an average between the two strong-line calibrators, which is the value adopted in the following figures. 
Errors on the adopted strong-line O/H take into account the flux uncertainties and the intrinsic errors of the method (0.18 and 0.16 dex, for O3N2 and N2, respectively, as quoted in M13). 
 
Comparing the average oxygen abundance derived with the \te-method, 
12+$\log$(O/H)=8.13$\pm$0.18, with the average values determined with the strong-line method, we have: for the N2 index 12+$\log$(O/H)=8.14$\pm$0.12, for the O3N2
index 12+$\log$(O/H)=8.20$\pm$0.14, and for the combination of the two indices 12+$\log$(O/H)=8.17$\pm$0.13. 
They are thus in extremely good agreement.
The increment of the number of regions analysed with the strong-line methods provides an even smaller dispersion of the distribution 
of the abundances in NGC~55 pointing towards a very homogeneous composition for the  interstellar medium for this galaxy.

\begin{table}
\centering
\begin{minipage}{75mm}
{\scriptsize  
\caption{Strong-line calibrated oxygen abundances.}
\begin{tabular}{@{}llllllll@{}}
\hline
Field-ID & D &O/H  &O/H   &  O/H &O/H \\         
 &kpc &T$_e$ &N2& O3N2 &  adopted \\          
 & &   &M13 & M13  & \\          
\hline    
H-1   &  12.13$^{+0.45}_{-1.35}$ & 8.21$\pm$0.07 & 8.03 & 7.97 &  8.03$\pm$0.21\\        
H-2   &  10.90$^{+0.14}_{-0.45}$ &8.15$\pm$0.06 & 8.14 & 8.18 &  8.16$\pm$0.21\\	  
H-3   & 10.84$^{+0.27}_{-0.84}$  &-	    & 8.18 & 8.27 &  8.22$\pm$0.30\\	  
H-4   &  10.21$^{+0.00}_{-0.00}$ & -        & 8.30  & 8.57  & 8.43$\pm$0.21\\   
H-5   & 9.87$^{+0.10}_{-0.32}$  & 8.23$\pm$0.06  & 8.10 & 8.13     & 8.12$\pm$0.21\\	  
H-6   & 9.01$^{+0.00}_{-0.01}$ & 7.90$\pm$0.10  & 8.25 & 8.32  & 8.28$\pm$0.22\\	  
H-7   & 9.43$^{+0.09}_{-0.29}$  & 7.97$\pm$0.12  & 8.11 &8.13 &  8.12$\pm$0.25\\	  
H-8   & 9.01$^{+0.05}_{-0.17}$ & -	    & 8.26 & 8.37&  8.32$\pm$0.21\\	  
H-9   & 8.42$^{+0.00}_{-0.02}$ &- & 8.12 & 8.14  & 8.13$\pm$0.23\\		  
H-10  & 9.52$^{+0.03}_{-0.11}$ &  - & 8.27 & 8.44 & 8.31$\pm$0.21\\		  
H-11  & 8.93$^{+0.07}_{-0.23}$ &  8.42$\pm$0.12  & 8.12 & 8.14 & 8.13$\pm$0.21\\	  
H-12  & 8.55$^{+0.19}_{-0.59}$ & 8.12$\pm$0.08 & 8.01  & 8.09 & 8.05$\pm$0.21\\ 	  
\\    
H-13  & 6.21$^{+0.06}_{-0.21}$  & $>$7.95 & 8.00 & 8.08 &  8.04$\pm$0.21\\      
H-14  & 6.46$^{+0.50}_{-1.41}$  & 8.06$\pm$0.07  & 8.07 & 8.11 &  8.09$\pm$0.22\\	 
H-15  & 5.77$^{+0.08}_{-0.24}$  &8.36$\pm$0.12       &-&- &-\\ 			 
H-16 & 5.40$^{+0.02}_{-0.07}$    &- & 8.22 & 8.21 & 8.21$\pm$0.21\\		 
H-17   & 4.42$^{+0.00}_{-0.01}$ & 8.20$\pm$0.07       &  8.12 & 8.18  & 8.15$\pm$0.22\\ 
H-18  & 7.63$^{+1.21}_{-3.03}$ &$>$8.25  & -&-&-   \\			 
\\
H-19 &  1.46$^{+0.18}_{-0.46}$  &7.70$\pm$0.12  & 7.83& 7.99  &7.91$\pm$0.21 \\    
H-20 & 0.95$^{+0.06}_{-0.18}$ & 7.86$\pm$0.13  &  8.10 & 8.16 &8.13$\pm$0.22\\      
H-21 & 2.12$^{+0.52}_{-1.17}$ & 7.80$\pm$0.12  & -&-&-  \\			     
H-22 & 1.00$^{+0.25}_{-0.56}$& 8.16$\pm$0.08 & 8.13 & 8.13 & 8.13$\pm$0.21\\     
H-23 & 0.28$^{+0.08}_{-0.16}$ &8.25$\pm$0.08  & 8.00 & 8.09 & 8.05$\pm$0.21 \\     
H-24 & 0.43$^{+0.01}_{-0.03}$ & - & 8.27 & 8.25 & 8.26$\pm$0.22\\	     
H-25 &0.98$^{+0.00}_{-0.00}$ & 8.38$\pm$0.12  & 8.05 & 8.05 &8.05$\pm$0.21\\	     
\hline					  
\end{tabular}
\label{tab_strong}
}
\end{minipage}
\end{table}

\section[]{Radial abundance gradients in NGC~55}
\label{sec_grad}

\begin{figure*} 
   \centering
   \includegraphics[width=18truecm]{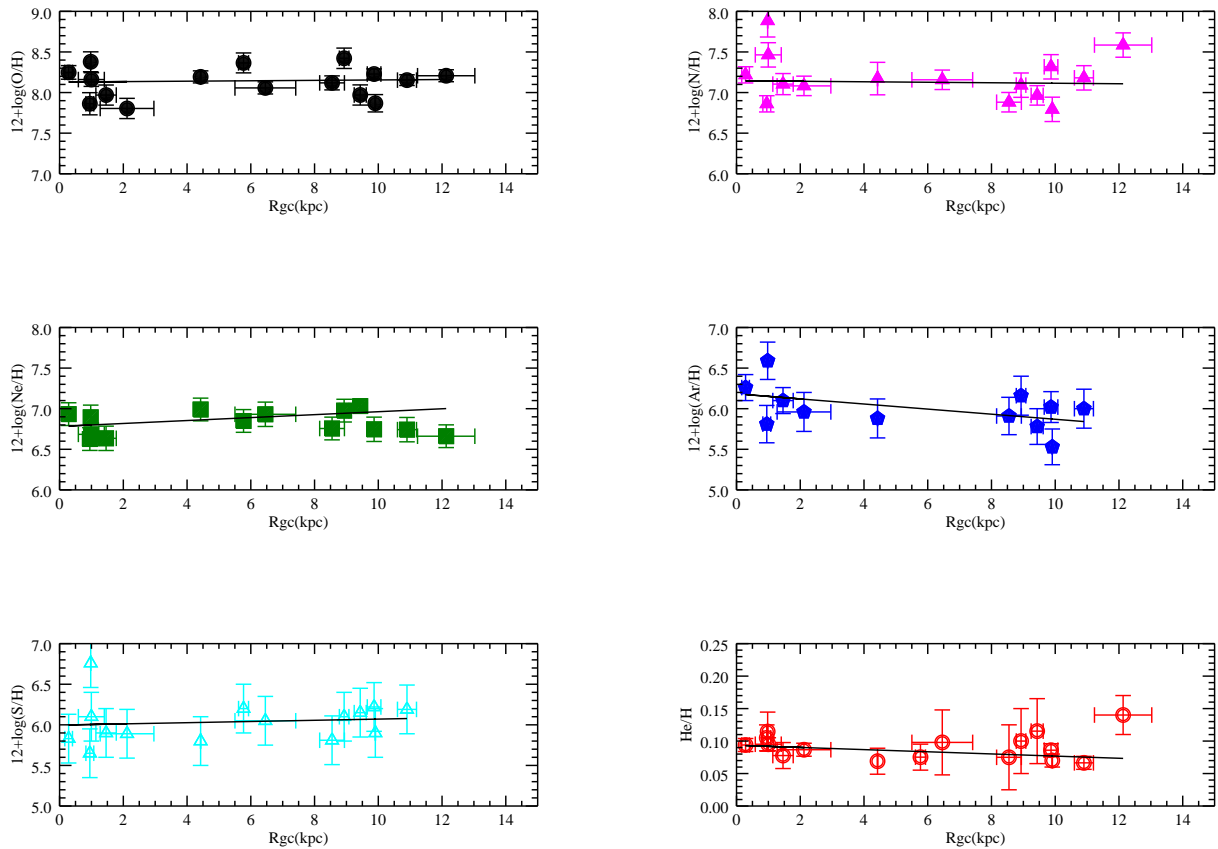} 
  \caption{Radial abundance gradients of elemental abundances in the  \hii\ regions in NGC~55. The abundances of metals (O, N, Ne, Ar, S) are expressed 
  in the logarithmic form 12$+\log$(O/H), whereas the abundance of helium is expressed in linear form.   
  The continuous curves are the weighted linear fits to the data,  taking into account errors on both distances and abundances.
    }
   \label{fig_grad_all}
\end{figure*}

\begin{table}
\caption{Radial abundance gradients in the disc of NGC~55}
\begin{tabular}{lll} 
\hline
El. & Slope & Intercept \\
\hline
O/H & +0.0025$\pm$0.0055 & 8.13$\pm$0.04\\
Ne/H & +0.0181$\pm$0.0081 & 6.78$\pm$0.06\\
N/H   & -0.0030$\pm$ 0.0082  & 7.14$\pm$0.06\\
Ar/H  & -0.0313$\pm$0.0145 & 6.18$\pm$0.10\\
S/H &  +0.0074$\pm$0.0196 & 5.99$\pm$0.13\\
He/H & -0.0017$\pm$0.0008 & 0.094$\pm$0.006\\
\hline
 N/O &-0.0103$\pm$0.0103 & -0.880$\pm$0.074\\
\hline
\end{tabular}
\label{tab_grad}
\end{table}

The \hii\ regions for which we can determine plasma conditions, including \te, \ne, ionic and total abundances, 
give us the opportunity to study the spatial distribution of abundances and abundance ratios of several elements in the thin/thick disc of NGC~55.  
We have computed the linear galactocentric distances de-projecting and transforming them in linear distances with the inclination, position angle and distance of Table \ref{tabNGC55}. 
We use the range of 4$^\circ$ in the inclination angle obtained by the disc model of  \citet{puche91} to estimate the uncertainties in de-projected galactocentric
distance.  The new model of \citet{westmeier13} is consistent with the one of Puche in the inner part of the galaxy  where our regions are located. 

The sample of \hii\ regions (see green symbols in Figure~\ref{fig_allsources}) are located in a large galactocentric range of distances (from the centre to about 12~kpc, see Table~\ref{tab_strong}). 

In Figure~\ref{fig_grad_all}  we show the radial abundance gradients of several elements:  O, Ne, S, Ar, N and He.  
In Table~\ref{tab_grad} we report the slopes and the intercepts of the radial abundance gradients  computed with the {\sc fitexy} routine that takes into account both the errors on abundances and galactocentric distances. For all the available elements we found null gradients, within the errors.


NGC~55 radial gradients of O/H and N/H have been also recently re-analysed from literature data by \citet{Pilyugin14}. 
They found  oxygen and nitrogen gradients essentially flat, and in good agreement with our results  (see Table\ref{tab_grad}). Their  slopes  for oxygen  and nitrogen are -0.0059$\pm$0.0104 dex~kpc$^{-1}$ and  -0.0042$\pm$ 0.0145  dex~kpc$^{-1}$, respectively. 
On the other hand, \citet{k16} analysed a sample of 58 blue supergiant stars. For the first time, they detected  a non negligible metallicity gradient of -0.22$\pm$0.06~dex/R$_{25}$ (-0.020$\pm$0.005 dex~kpc$^{-1}$ assuming R$_{25}$=11.0~kpc). Their central  metallicity  relative to the Sun is Z=-0.37$\pm$0.03.

\begin{figure} 
   \centering
   \includegraphics[width=9truecm]{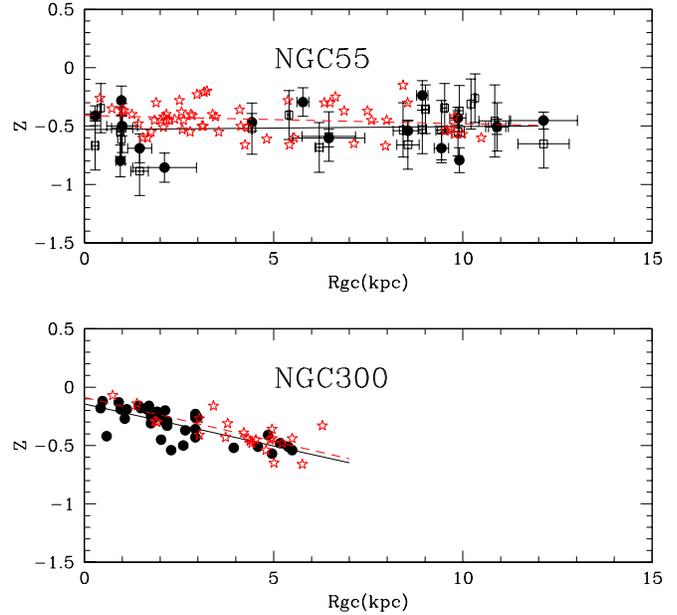} 
  \caption{Upper panel: NGC~55 \hii\ regions  (O/H derived with the \te\ method --filled circles-- and from the strong-line method --empty squares. For the strong-line abundances we adopted the average of the two values, as indicated in Table~\ref{tab_strong}) and supergiants from \citet{k16} (empty stars) radial distribution. 
  The continuous line is the fit to the \hii\ regions with \te\ determinations as in Figure \ref{fig_grad_all}, whereas the dotted line is the fit to the whole sample of supergiant abundances. 
 Lower panel: NGC~300 \hii\ regions  \citep[filled circles from]{Bresolin09, Stasinska13} and supergiants from \citet{k08} (empty stars) radial distribution.   
 }
   \label{fig_gas_stars}
\end{figure}

The comparison between the \hii\ region and blue supergiant populations is extremely interesting because they both are young populations and should trace the same epoch in the galaxy lifetime.  
In the upper panel of Figure~\ref{fig_gas_stars} we plotted the metallicity of the supergiants of \citet{k16} and of our \hii\ regions --reported on the Solar scale \citep[12+$\log$(O/H)=8.66;][]{grevesse07}-- versus the galactocentric distance. We have plotted the whole sample 
of \citet{k16} without removing the possible outliers. 
The two sets of metallicities and their radial distributions are in surprisingly good agreement considering all the internal uncertainties of the two metallicity derivations, the differences in the metallicities measured from nebular and stellar spectra (oxygen in the former, a global Z metallicity in the latter), and possible dust depletion of oxygen in \hii\ regions \citep[this correction might amount to 0.10~dex
for objects with 7.8$<$12+$\log$(O/H)$<$ 8.3); see ][]{pp10}.
It is, however, true that the supergiant abundances show a small decreasing gradient, which is not appreciable in the \hii\ region population abundances.

\subsection{$\alpha$-elements and nitrogen}
Giving the similar origin of the four $\alpha$-elements --O, Ne, S and Ar--  we expect them to have similar radial behaviours. We indeed find almost flat gradients within the uncertainties for all of them, with 
Ar  having possibly a small negative slope, still consistent with a null gradient,  within the errors. 

While $\alpha$-elements are mainly synthesised by massive stars (M$>$8~M$_{\odot}$), nitrogen has a more complex nucleo-genesis, 
having both a \lq\lq primary" and a \lq\lq secondary" origin. The primary origin refers to conversion of the original hydrogen into nitrogen and it happens in stars with 4~M$_{\odot}<$M$<$8~M$_{\odot}$
\citep{RV81} and/or in very massive (M$>$30M$_{\odot}$) low metallicity stars \citep{WW95}, while the secondary channel 
is related with the production from C and O initially present in the ISM at the formation of the progenitor star. 
When the primary nitrogen component dominates, the N/O ratio is expected to be independent of the oxygen abundance and this happens at low metallicity. 
At higher metallicity, the secondary production becomes more important, and N/O increases with O \citep{vZ98}.

\begin{figure} 
   \centering
   \includegraphics[width=9truecm]{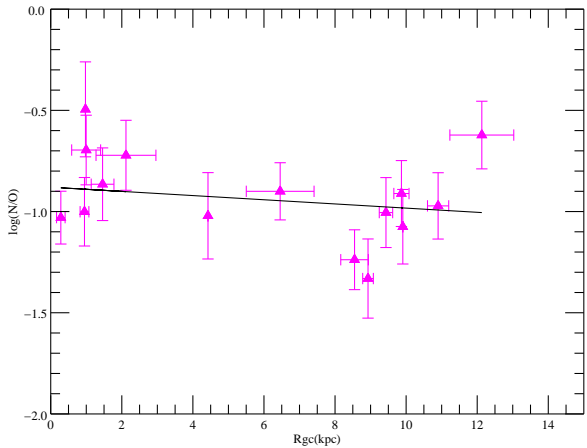} 
  \caption{Radial N/O gradients of \hii\ regions of NGC~55.    }
   \label{fig_grad_no}
\end{figure}

The comparison of N with any $\alpha$-element can give indication on different  star formation history in different radial regions of NGC~55 disc.  
If a galaxy experiences a dominant global burst of star formation, the ISM oxygen abundance will increase in about 10$^6$~yr, generating 
a decrease in N/O. Then over a period of several 100$\times$10$^6$~yr N/O will increase at constant O/H. 
Consequently N/O can be used as  a clock that measures the time since the last major burst of star formation: low values of N/O imply a very recent burst of star formation, while high values of N/O imply a long quiescent period  \citep[cf.][]{skillman98}.
In Figure~\ref{fig_grad_no}, the radial gradient of N/O is shown. A slightly decreasing gradient towards the outskirts is detected. 
However this gradient is consistent with 2-$\sigma$ with a flat gradient and homogeneous distribution of abundances, indicating no differences in the star formation histories in different parts of 
the galaxy.  
We can also compare the typical N/O value in NGC~55 with those of dwarf and spiral star forming galaxies. 
From Figure~5 of \citet{Pilyugin03} or Figure~7 of \citet{Annibali15} we can infer that at 12$+\log$O/H$\sim$8.0~dex, there is an increasing dispersion in N/O, with values 
ranging from  $-$1.6 to $-1$~dex. 
N/O in NGC~55 is located towards the upper envelope of this relationship.  The average  N/O of NGC~55 (N/O$=-$0.93$\pm$0.21) is close to the M33's one 
(see, e.g., Figure 12 in Magrini et al. 2007,  
for a plot of N/O vs O/H) and to that of the LMC \citep[N/O=-0.96;][]{CR15}.
The scatter in N/O values at a given O/H seen, e.g., by \citet{Pilyugin03} and  \citet{Annibali15}, can be naturally explained by differences
in the star formation histories of galaxies \citep{Pilyugin03}. This conclusion was already  suggested by \citet{EP78} who explained the  observations of the N/O abundance ratio in
external galaxies due to the manufacturing of N in low-mass stars of 4-8 M$_{\odot}$. 
Very recently, \citet{vincenzo16}  presented chemical evolution models aimed at reproducing the observed N/O versus O/H abundance pattern of star-forming galaxies in the Local Universe.
They found that position of a galaxy in the N/O vs. O/H plane is mostly determined by its star formation efficiency (see their Figure 4). 
The high N/O ratio in NGC~55 with respect to other late-type dwarf irregular galaxies having the same O/H might  
be an indication  that the bulk of the star formation happened in the recent past, more than 100$\times$10$^6$~yr ago. 
This is in agreement with the finding of \citet{Davidge05} of a vigorous star formation episode during the past 0.1-0.2 Gyr. 
The detection of significant numbers of stars evolving on the AGB phase, indicates that there has been vigorous star formation during
the past 0.1-0.2 Gyr. In this time lapse, stars with masses  M$>$5~M$_{\odot}$ might had time to evolve and to pollute with N the ISM of NGC~55 and thus to increase the N/O ratio. 

\subsection{Helium}
The measurement of He in low metallicity galaxies has been often used to extrapolate the primordial He abundance. 
To account for the amount of helium synthesised by stars, \citet{ptp74} suggested for the first time to study a sample of \hii\ regions spanning 
a wide range of metallicity. With the assumption of a helium enrichment proportional to metallicity, we can obtain the primordial abundance of 
He extrapolating to zero metallicity: $$Y=Y_{\rm P}+Z(\Delta Y/\Delta Z),$$ where Y is the mass fraction of He, and Z is the metallicity. 
Using Equation~2 of \citet{IT97} without correction for the neutral He \citep[$<$2\%, cf.][]{IT97}, we computed the mass fraction of He using the mean abundance of NGC~55, obtaining Y=0.27$\pm$0.08. Excluding the \hii\ region with the highest He abundance (H-1) we have a mean He/H=0.088$\pm$0.020 and thus Y=0.26$\pm$0.04. 
The linear regression of \citet{IT97} of Y versus oxygen, 
computed at the metallicity of NGC~55 gives a slightly lower value Y=0.250$\pm$0.002, but still consistent within the errors 
with our determination. 
The measurement of He abundance in NGC~55 does not give any particular constraint to the primordial He abundance, but it is in line with the determination of Y in several low metallicity galaxies. 
 
 \section{Discussion}
 \label{sec_dis}

NGC~55 has both been classified as an irregular and a spiral galaxy. However, although its morphology cannot be well defined because of its inclination, 
it is a very extended object and with a large total mass (see Table~\ref{tabNGC55}); 
its \hii\ regions are found up to 11-12~kpc from the centre, and its blue and red supergiants even much farther \citep{k16}. 
So, the questions are: how can the composition of its \hii\ regions (and supergiants) be so homogeneous in such a wide region, extending from the centre to about $\sim$11-12~kpc? 
Why is not there a clear metallicity gradient from the \hii\ region abundances? 

The presence of metallicity gradients in irregular galaxies has been, indeed, deeply investigated in the literature reaching the conclusion that  
most of these galaxies show a spatial homogenous composition  \citep[e.g.,][]{KS97, Croxall09, MG09, Haurberg13, LP13, hosek14, patrik15}. There are some exceptions, such as, for instance, the observed oxygen gradient in NGC 6822 \citep{venn04, Lee06} and in the dwarf blue compact galaxy NGC1750 \citep{Annibali15}, or the iron gradients in the SMC, in the LMC, and in  the dIrr WLM \citep{Leaman14}.  
The systematic study of gradients in late-type spiral and irregular galaxies  by \citet{Pilyugin14, Pilyugin15} has shown that there is a strong correlation 
between the radial abundance gradient of galaxies and their surface brightness
profiles. In particular, galaxies with a steep inner surface brightness profile usually 
present a noticeable radial abundance
gradient. On the other hand, those galaxies  with flat inner surface brightness
profiles have shallower or null gradients. 
The reason of this different behaviour might be related to the presence of  radial mixing of gas (through
radial flows or galactic fountains): this process takes place more evidently in  galaxies
with a flat surface density inner profile that can readily redistribute the gas across the disc. 

The comparison of metallicity distributions and surface density profiles in the galaxy pair, NGC~55 and NGC~300 (Scd for NGC~300 and barred spiral/irregular SB(s)m for NGC~55) allows us to investigate this hypothesis in detail. 
These two galaxies are indeed very similar in many aspects: comparable near-IR magnitudes \citep[K=6.25 and 6.38, respectively, see][]{Jarrett03} 
and mid-infrared fluxes \citep[see][]{Dale09} that point towards comparable stellar masses. 
However, they have different morphologies. 
Even more important, \citet{k16} have shown that their K-band radial surface profiles are also different, with NGC~300 having a strong central peak, which is instead absent in NGC~55 (see their Figure~12). 
The radial metallicity gradient of NGC~300 from \hii\ regions, studied by \citet{Bresolin09} and re-analysed by \citet{Stasinska13},  
is not negligible, and has a slope -0.068$\pm$0.009~dex~kpc$^{-1}$ (with R$_{25}$=5.3~kpc),  as well as the radial metallicity gradient 
of supergiants \citep{k08, gazak15}.
These gradients are shown in the lower panel of Figure~\ref{fig_gas_stars}. 

Thus, in the light of the discussion of \citet{Pilyugin15}, it seems that the different K-band radial surface profiles imply 
a different mass accretion/gas outflow in the two galaxies, with a consequent much higher re-distribution of metals in NGC~55 with respect to NGC~300. 
\citet{k16} concluded that the chemical evolution of NGC~55
is characterised by large amounts of gas outflow and infall, having one of the largest known 
 infall rate in the Local Universe. This agrees well with the detailed morphological,
kinematic and dynamic study of the neutral gas by \citet{westmeier13}  who concluded that internal and external processes,
such as satellite accretion or gas outflow  have stirred up the gas disc, inducing 
streaming motions of the gas along the bar of NGC~55. In addition, \citet{westmeier13} made evident several 
isolated \hi\ clouds within about 20 kpc projected distance from NGC~55. 
These clouds are similar  to the high-velocity
clouds of the Milky Way and they are a signature of on-going infall of gas and of the complex dynamics of NGC~55. 

We also took advantage of the available X-rays data for NGC~55 to test the possibility that the intense star-forming events of the galaxy blow out the heavy elements, out of its disc to the galaxy halo. As detailed in Appendix~A, the X-rays luminosity and SFR from the 12 of our 25 \hii\ regions for which data exists in the Chandra archive, are not consistent with the scenario on which the flat metallicity gradient would be justified by blow up of heavy elements due to extreme winds. \citet{stobbart06} (their sources that are located within the optical confines of the galaxy are signed in Figure~\ref{fig_allsources}), studying the X-ray properties of NGC~55 using XMM-Newton observations, derived the physical properties of the soft residual disc component and found the SFR equal to 0.22 M$_\odot$ yr$^{-1}$ and the pressure of the hot gas (P/$k$) equal to $\sim$ 1.5$\times$10$^5$ K cm$^{-3}$. This later value is similar to the pressure inferred for the interior of the Loop 1 superbubble within our Galaxy \citep{2003Willingale}. The authors conclude that although this is broadly consistent with the scenario where there has been sufficient recent star formation in the disc of NGC~55 to form expanding bubbles responsible for ejecting the material out of the disc into the halo, the absence of an extended extra-planar soft X-ray component in NGC~55 \citep[contrary to what was found by][]{2002Oshima} points to the fact that the gas in such bubbles cools relatively quickly through adiabatic losses, retaining insufficient energy to power a superwind of the form frequently seen in systems with SFR $>$ 1 M$_\odot$ yr$^{-1}$ \citep{strickland04}. So, our \hii\ region sample does not produce enough superwind in NGC~55, which is in agreement with previous analysis, though we should keep in mind that from all the star formation indicators, the X-ray is the weakest one. 

Therefore all the above discussions suggest that the dominant effects in flattening the metallicity gradient of NGC~55 are those related to the dynamics of the gas through bar-driven mixing and inflow/outflow. 

\section{Conclusions}
\label{sec_conclu}

In the present paper we show new spectroscopic observations of a large sample of 25 \hii\ regions in the Sculptor group member galaxy, NGC~55. 
We derive physical and chemical properties though the \te-method of 18 \hii\ regions, and strong-line abundances for 22 \hii\ regions. 
We measure also abundances of He, O, N, Ne, S, Ar. We found a homogenous composition of the disc of NGC~55, with average abundances of  He/H=0.092$\pm$0.019, 12+$\log$(O/H)=8.13$\pm$0.18, 12+$\log$(N/H)=7.18$\pm$0.28, 12+$\log$(Ne/H)=6.81$\pm$0.14, 12+$\log$(Ar/H)=6.00$\pm$0.25 and 12+$\log$(S/H)=6.04$\pm$0.25. The abundances are uniformly distributed in the radial  direction. 
This agrees with the study of smaller samples of \hii\ regions \citep{WS83, Pilyugin14} and it is in qualitative agreement with the blue supergiant radial gradient \citep{k16}. We investigate the origin of such flat gradient comparing NGC~55 with its companion galaxy, NGC~300, similar in terms of mass and luminosity and located in the same group of galaxies. The most plausible hypothesis is related to the differences in their K-band surface density profile that, as suggested by \citet{Pilyugin15}, which can provide higher mixing of the NGC55 gaseous component than in similar galaxies.

\section{Acknowledgments}
We are extremely grateful to Rolf Kudritzki for his constructive and helpful report. 
The work of DRG was partially supported by FAPERJ's grant APQ5-210.014/2016.  The work of Bruna Vajgel is sponsored by grants from CNPq, process number 164858/2015-6.
This research has made use of data obtained from the Chandra Data Archive and software provided by the Chandra X-ray Center (CXC) in the application packages CIAO.

{}


%
\appendix
\section[]{X-ray Luminosity}

We complemented our observations with the available X-ray observations of the \hii\ regions in NGC~55 in the Chandra archive. 
\par The analysis of the X-ray emission for the 25 nebulae consists basically of measuring the source net counts and then converting the count rate into X-ray flux and luminosity. We used archival Chandra observation (ObsID 2255) with exposure of 60.1 ks of NGC~55, performed in 2001 September with the Advanced CCD Imaging Spectrometer (ACIS-I). The Chandra data were reduced and reprocessed using the science threads of Chandra Interactive Analysis of Observations (CIAO) version 4.6.

\begin{table}
\centering
\begin{minipage}{70mm}
{\scriptsize  
\caption{X-ray net counts, luminosities and SFR of the \ha\ line-emitters from Table 1.}
\begin{tabular}{@{}lllll@{}}
\hline
Field-ID &Net Counts &Aperture  &L$_{\rm 2-10keV}$       &SFR \\
         &(photons)  &(``)      &(10$^{36}$ erg s$^{-1}$) &(10$^{-3}$ M$_{\odot}$ yr$^{-1}$)  \\        
\hline    
H-1&           -0.33&  1.5&      $<$1.53&         $<$0.40 \\
H-2&	       -2.36&  4.0&	  $<$10.94&        $<$2.88 \\	  
H-3&	       -0.60&  2.0&	  $<$2.79&         $<$0.73 \\ 		  
H-4&	        0.67&  1.5&	  3.11$\pm$4.64&   0.82 \\
H-5&	       -0.92&  2.5&	  $<$4.26&         $<$1.12 \\		  
H-6&	       -0.32&  3.0&	  $<$4.64&         $<$1.22 \\ 		  
H-7&	        1.69&  3.0&	  7.82$\pm$8.05&   2.06 \\
H-8&	       -1.32&  3.0&	  $<$6.10&         $<$1.61 \\ 		  
H-9&	        2.78&  7.0&	  12.89$\pm$14.84& 3.39 \\  
H-10&	       -0.59&  2.0&	  $<$2.74&         $<$0.72 \\ 		  
H-11&	        0.22&  3.5&	  1.02$\pm$6.58&   0.27 \\
H-12&	       -2.02&  4.5&	  $<$4.64&         $<$1.22 \\ 
\\		  
H-13&	        6.69&  6.0&	  31.06$\pm$16.16& 8.17 \\	  
H-14&	       -0.33&  3.0&	  $<$4.64&         $<$1.22 \\		  
H-15&	       -0.95&  4.5&	  $<$9.28&         $<$2.44 \\		  
H-16&	        0.67&  3.0&	  3.12$\pm$6.58&   0.82 \\  
H-17&	        4.65&  6.0&	  21.57$\pm$14.77& 5.68 \\	  
H-18&	        0.40&  2.0&	  1.85$\pm$4.65&   0.49 \\ 
\\
H-19&	       -0.74&  10.0&	  $<$64.99&        $<$17.10 \\		  
H-20&	       -0.35&  3.0&	  $<$4.64&         $<$1.22 \\ 		  
H-21&	        1.41&  2.0&	  6.54$\pm$6.57&   1.72 \\
H-22&	       -0.35&  4.0&	  $<$9.28&         $<$0.002 \\ 		  
H-23&	        2.64&  4.0&	  12.26$\pm$10.40& 3.23 \\	  
H-24&	        0.07&  2.5&	  0.32$\pm$4.65&   0.09 \\  
H-25&	        1.09&  2.5&	  5.07$\pm$6.57&   1.33 \\  
\hline					  
\end{tabular}
}
\end{minipage}
\label{tab_lumxray}
\end{table}

The first step after reprocessing the data has been to estimate the background counts used to obtain the net source counts for each H$\alpha$ emitter from Table 1.  The background contribution in the 2-10 keV band has been evaluated in a nearby source-free circular region with a radius of 30\arcsec. The X-ray count rates of the nebulae were estimated inside a circular region too. The optimal size of each H$\alpha$ emitter has been measured by eye in the H$\alpha$ images. The background count normalised by the source area has been subtracted from the source count. The net count rates were obtained by dividing the net counts by the data exposure time.  To convert the net count rates to X-ray flux in the 2-10 keV energy band, we use the \texttt{PIMMS}\footnote{Available on the HEASARC-NASA website} software package routine. We calculated the count-to-energy conversion assuming a given spectral model, temperature, abundance and hydrogen column density.  We adopted the astrophysical plasma emission code mekal \citep{liedahl95}, with a metallicity equal to 0.4 $Z_\odot$ and  a temperature of 1~keV.  The hydrogen column density (21~cm) towards NGC55  was obtained from Leiden/Argentine/Bonn (LAB) Survey of Galactic H~{\sc i} and is equal to 1.37$\times 10^{20}$ cm$^{-2}$.  Once the net count rates were converted to fluxes, we determined the luminosity, L$_{2-10keV}$. The $k$-correction from \citet{bohringer00} has been applied to obtain the rest-frame X-ray luminosity. From the rest-frame L$_{2-10keV}$ we estimated the star formation rate using:

\begin{equation}
{\rm SFR}(>0.1 {\rm M_{\odot}}) = \frac{L_{\rm 2-10keV}}{3.8\times 10^{39} {\rm erg ~s^{-1}}} {\rm M_{\odot}~ yr^{-1}},
\end{equation}

\noindent from \citet{PR07}. Because for most of the sources we have only X-ray upper limits, i.e. their emission is below the background level, we did not attempt to perform any quantitative analysis in this energy band. This quantities are only used to have an insight of the SFR. The so-obtained net counts, used apertures to measure the X-ray emission, luminosities and the respectively SFR are listed in Table \ref{tab_lumxray}. The negative values of the net counts mean that their emission is below the background level, these correspond to the sources with X-ray upper limit. From the 25 H$\alpha$ emitters only 12 have reliable measured X-ray emission. The rest-frame X-ray luminosities span from 10$^{35}$ to 10$^{37}$~erg~s$^{-1}$ and their respective SFRs range from 10$^{-4}$ to 10$^{-3}$~M$_{\odot}$~yr$^{-1}$. The SFR of the individual H$\alpha$ emitters is, as expected,  much lower
than the global SFR of NGC55 \citep[$\sim$0.22~M$_{\odot}$~yr$^{-1}$;][]{eng04} estimated with \textit{Spitzer} far-infrared (24~$\mu$m) data.

\section{Emission-line flux measurements}
In this section we present the observed emission-line fluxes and extinction corrected intensities 
measured in our \hii\ regions. 

\begin{table}
\centering
\begin{minipage}{82mm}
{\tiny  
\caption{Observed fluxes and extinction corrected intensities. Column
(1) gives the target name; column (2) gives the observed \hb\ flux (with error)
in units of 10$^{-16}$~erg cm$^{-2}$ s$^{-1}$; 
column (3) the nebular extinction coefficient (with error);
 columns (4) and (5) indicate the emitting ion and the rest frame wavelength
in \AA; columns (6), (7), and (8) give extinction corrected (I$_{\lambda}$) intensities, the 
relative error on the fluxes ($\Delta$F$_{\lambda}$) and the measured fluxes (F$_{\lambda}$). Both I$_{\lambda}$ and F$_{\lambda}$ are normalised to \hb=100. }
\begin{tabular}{@{}ccclclll@{}}
\hline
Id & F$_{{\rm H}\beta}$ & \cbeta\         & Ion & $\lambda$ (\AA) & I$_{\lambda}$ & $\Delta$F$_{\lambda}$  & F$_{\lambda}$ \\ 
   &     $\Delta$F$_{\lambda}$               & $\Delta$\cbeta\ &     &                 &               &        (\%)   	     &  	     \\ 
\hline
H-1&3.6e-16& 0.00         & [O~{\sc ii}]   &  3727  &    9.9 & 0.5 &    9.9	 \\ 
    & $\pm$2e-17& $\pm$0.1& H~{\sc i}&  3798  &   3.7  &  0.5 &   3.7	\\
       &      &   	     & [Ne~{\sc iii}] &  3868  &  10.   &  0.5 &  10.  \\
       &      &   	     & [Ne~{\sc iii}] &  3968  &  11.  & 0.5 & 11.   \\
       &      &   	     & H$\delta$      &  4100  &  8.8   & 0.5 &  8.8 	\\
       &      &   	     & H$\gamma$      &  4340  & 36.  & 2. &  36.	\\
       &      &   	     & [O~{\sc iii}]  &  4363  &  7.5  & 0.5 &  7.5    \\
       &      &   	     & He~{\sc i}     &  4473  &   5.8  & 0.5 &   5.8	\\
       &      &   	     &  O~{\sc ii}    &  4671	&  44.  & 2. &  44. \\
       &      &   	     &  N~{\sc ii}    &  4788	&  27.  & 2. &  27. \\
       &      &   	     & H$\beta$       &  4861   & 100.  & 1. & 100.	\\
        &      &   	     & [O~{\sc iii}]  &  4959  & 239.  &15. & 239.    \\
       &      &   	     & [O~{\sc iii}]  &  5007  & 662. &30. &662.  \\
        &      &   	     & He~{\sc i}     &  5876  &  19.  &2. &  19.	\\
       &      &   	     & H$\alpha$      &  6563  & 260.  & 15. & 260.	\\
       &      &   	     & [N~{\sc ii}]   &  6584  &  6.8   & 0.5  &  6.8 	\\
       &      &   	     & He~{\sc i}     &  7065  &  11.5  & 1.  &  11.5	\\
\hline      
H-2 &1.2e-15&0.026      & 	[O~{\sc ii}]   &  3727  &  22. &  1. &  21. 	\\    
     &6e-17&$\pm$0.05&  [Ne~{\sc iii}] &  3868  &  3.2 &  1. &   3.2     \\
      &      &   	   & H~{\sc i}      &  3889  &  4.5  &  0.5 &   4.4	\\
     &      &   	   & [Ne~{\sc iii}] &  3968  &  5.7  &  0.5 &  5.6     \\
     &      &   	   & H$\delta$      &  4100  &  10.  &  0.5 &  10.	 \\
     &      &              & H$\gamma$      &  4340  &  32.  &  2. &  32.	 \\
     &      &   	   & [O~{\sc iii}]  &  4363  &   1.1  &  0.5 &   1.1	 \\
     &      &   	   & He~{\sc i}     &  4471  &   2.6  &  0.5 &   2.6	\\
     &      &   	   & H$\beta$	    &  4861  & 100.  &   5. & 100.   \\
     &      &   	   & [O~{\sc iii}]  &  4959  &  75.  &   4. &  75.   \\
     &      &   	   & [O~{\sc iii}]  &  5007  & 235.  &  12. & 235. \\
     &      &   	   & He~{\sc i}     &  5876  &  9.4  &  1. &  9.4   \\
     &      &		   & H$\alpha$	   &  6563  & 287.  &  15. & 292.	 \\
     &      &		   & [N~{\sc ii}]   &  6584  &  14.  &   1. &  14.	 \\
     &      &		   & He~{\sc i}     &  6678  &	3.  &   1. &   3.	 \\
     &      &		   & [S~{\sc ii}]   &  6717  &  19.6  &   1. &  20.0	 \\
     &      &		   & [S~{\sc ii}]   &  6731  &  12.3  &   1. &  12.6	 \\
     &      &		   & He~{\sc i}     &  7065  &	2.7  &   0.5 &   2.8	 \\
     &      &		   & [Ar~{\sc iii}] &  7135  &	8.3  &   0.5 &   8.5	 \\
     &      &		   & [O~{\sc ii}]   &  7320  &	3.3  &   0.5 &   3.4	 \\
     &      &		   & [O~{\sc ii}]   &  7330  &	1.6  &   0.5 &   1.7	 \\
     &      &		   &He~{\sc i}	    &  7816  &	2.4  &   0.5 &   2.5	 \\
\hline   	    
H-3 &1.4e-16& 2.0 &  H~{\sc i}    &  3721  &  30. & 1. &  7.8  \\
    &$\pm$1e-17&$\pm$0.2& [O~{\sc ii}]&  3727  &  286. & 4. &   75.  \\
     &      &   	   & H$\gamma$      &  4340  &  41. & 1. &  22.  \\
      &      &   	   & H$\beta$	    &  4861  &   100. & 5. & 100.  \\
     &      &   	   & [O~{\sc iii}]  &  4959  &   42. & 3. &  47.  \\
     &      &   	   & [O~{\sc iii}]  &  5007  &  109. & 6. & 130.  \\
     &      &		  & [N~{\sc ii}]   &  6548  &	9.  &   3. &  39.	\\
     &      &		  & H$\alpha$	   &  6563  & 287.  &  50. & 1270.  \\
     &      &		  & [N~{\sc ii}]   &  6584  &  18.  &   5. &  75.	\\
     &      &		 & [S~{\sc ii}]    &  6717  &  16.  &   5. &  76.	 \\
     &      &		 & [S~{\sc ii}]    &  6731  &  13.  &   4. &  61.	 \\
     &      &		  & [Ar~{\sc iii}] &  7135  &	4.  &   2. &  25.	\\
     &      &		  & [O~{\sc ii}]   &  7320  &	4.  &   2. &  28.	\\
     &      &		  & [O~{\sc ii}]   &  7330  &	3.  &   2. &  19.	\\
     &      &		  & [S~{\sc iii}]  &  9069  &	2.  &   2. &  44.	\\
\hline 
H-4&1.7e-16&0.46          &  He~{\sc i} &  3554  &  4.5 & 1. &  3.2  \\
    &$\pm$1e-17&$\pm$0.05& [O~{\sc ii}]&  3727  &  52. & 2. &   38.  \\
     &      &   	   & H$\gamma$      &  4340  &  38. & 2. &  33.  \\
     &      &   	   & H$\beta$	    &  4861  &  100.& 5. & 100.  \\
     &      &   	   & [O~{\sc iii}]  &  5007  &   7.0 & 1. & 7.3  \\
     &      &		  & H$\alpha$	   &  6563  &  287.  &  15. & 401.  \\
     &      &		  & [N~{\sc ii}]   &  6584  &   31.  &   3. &  43. \\
     &      &		 & [S~{\sc ii}]    &  6717  &   41.  &   4. &  59.   \\
     &      &		 & [S~{\sc ii}]    &  6731  &   31.  &   3. &  43.   \\
\hline
H-5&1.05e-15& 0.11        & [O~{\sc ii}]   &  3727  & 38. & 2. & 35. \\
 &$\pm$5.e-17&$\pm$0.05 & [Ne~{\sc iii}] &  3868  &  4. & 1. &  4. \\  
     &     &       	     &  H~{\sc i}     &  3889  &  4. & 1. &  4.   \\
       &     &		    & [Ne~{\sc iii}] &  3968  &   6. & 1. &   5.   \\
     &     &		    & H$\delta$      &  4100  &  11. & 1. &  11.   \\
     &     &		    & H$\gamma$      &  4340  &  32.  & 2. &  31.   \\
     &     &		    & [O~{\sc iii}]  &  4363  &   1.7 & 0.5 &   1.6   \\
     &     &		    & He~{\sc i}     &  4471  &   3.3 & 0.5 &   3.3   \\
     &     &		    &  S~{\sc ii}   &  4791  &   2.9   &  0.5 &  2.9   \\  			  
     &     &		    & H$\beta$       &  4861  & 100.	&  5.  & 100.   \\
     &     &		    & [O~{\sc iii}]  &  4959  & 108.	&  5.  & 109.   \\
     &     &		    & [O~{\sc iii}]  &  5007  & 334.	& 16.  & 337.   \\
      &     &		    & He~{\sc i}     &  5876  &  13.	&  1.  &  14.  \\
     &     &		    & H$\alpha$      &  6563  & 287.	&  15.  & 310. \\
     &     &		    & [N~{\sc ii}]   &  6584  &  12.	&  1.  &  13. \\
     &     &		    & He~{\sc i}     &  6678  &   3.2	&  1.  &   3.5\\
     &     &		    & [S~{\sc ii}]   &  6717  &  18.		&  1. &  19.\\
     &     &		    & [S~{\sc ii}]   &  6731  &  13.		&  1.  &  15. \\
     &     &		    & He~{\sc i}     &  7065  &   4.9	&  1.  &   5.4 \\   
     &     &		    & [Ar~{\sc iii}] &  7135  &   8.1	&  1.  &   8.9  \\  
\hline
\end{tabular}
}
\end{minipage}
\label{tabPN_flux}
\end{table}
\begin{table}
\centering
\begin{minipage}{82mm}
{\tiny
\contcaption{}
\begin{tabular}{@{}ccclclll@{}}
\hline
Id & F$_{{\rm H}\beta}$ & \cbeta\ & Ion & $\lambda$ (\AA) & I$_{\lambda}$ & $\Delta$F$_{\lambda}$ & F$_{\lambda}$ \\ 
\hline
H-6&6.7e-16& 0.68       & [O~{\sc ii}]   &  3727 & 52.    & 2.   &  33.\\
 &$\pm$3e-17&$\pm$0.05& [Ne~{\sc iii}] &3968 &  5.7    & 0.5   &   4.0   \\  
     &     &		    & H$\delta$      &  4100  &   9.2	&  0.5  &   6.8   \\
     &     &		    & H$\gamma$      &  4340  &  32.&  2.  &  26.   \\
     &     &		    & [O~{\sc iii}]  &  4363  &   0.8	&  0.3 &   0.7   \\
     &     &		    & H$\beta$       &  4861  &  100.	&  4.  & 100.   \\
     &     &		    & [O~{\sc iii}]  &  4959  &  29.	&  2.  &  30.   \\
     &     &		    & [O~{\sc iii}]  &  5007  &  87.	&  5.  &  92.   \\
     &     &		    & He~{\sc i}     &  5876  &   9.2	&  2.  &  13.  \\
     &     &                & [N~{\sc ii}]   &  6548  &  12.	&  1.  &  19.	 \\	
     &     &		    & H$\alpha$      &  6563  & 287.	&  10.  & 469. \\
     &     &		    & [N~{\sc ii}]   &  6584  &  24.	&  3.  &  39. \\
     &     &		    & [S~{\sc ii}]   &  6717  &  25.		&  3.  &  41. \\
     &     &		    & [S~{\sc ii}]   &  6731  &  18. 	&  2.  &  29. \\
     &     &		    & [Ar~{\sc iii}] &  7135  &   4. 		&  1.  &   8.  \\  
     &     &                & [O~{\sc ii}]   &  7320  &   4.	&  1.  &   7. 	\\
     &     &                & [O~{\sc ii}]   &  7330  &   2.	&  0.5  &   4.	\\
\hline
H-7&1.56e-15&1.5        & [O~{\sc ii}]   & 3727 & 75. & 2. &  29. \\
   &$\pm$8e-17&$\pm$0.1& H~{\sc i} & 3797 &  5.0 & 1. &   2.0 \\  
     &     &	    &	He~{\sc i}    & 3833 &  9.5 & 1. &   4.0 \\			 
     &     &	   &	 [Ne~{\sc iii}]      &  3868  &  20. &  0.5&   9.\\  
     &     &    &  H~{\sc i}	 &  3889  &  10.  &  0.5&   4.4   \\
     &     &    & [Ne~{\sc iii}] &  3968  &  26. &  1. &  12.   \\
     &     &    & He~{\sc i} & 4023   &   4.3 &  0.5&   2.1\\
      &     &    & H$\delta$	 &  4100  &  23. &  0.6&  12.   \\
     &     &    & H$\gamma$	 &  4340  &  47.  &  3. &  30.   \\
     &     &    & [O~{\sc iii}]  &  4363  &   4.3 &  0.5 &   2.8   \\
     &     &    & H$\beta$	 &  4861  & 100. &  4. & 100.   \\
     &     &    & [O~{\sc iii}]  &  4959  & 169. &  9. & 183.   \\
     &     &    & [O~{\sc iii}]  &  5007  & 362.  & 15. & 410.   \\
       &     &   & He~{\sc i}	&  5876  &  15.  &  2. &  30. \\
     &     &   & [N~{\sc ii}]	&  6548  &   3.  &  0.6 &   9.    \\   
     &     &   & H$\alpha$	&  6563  & 287.  & 20. & 840.    \\   
     &     &   & [N~{\sc ii}]	&  6584  &  12. &  2. &  36.    \\   
      &     &   & He~{\sc i}	&  6678  &   5. &  1. &  14.  \\    
     &     &   & [S~{\sc ii}]	&  6717  &  12.  &  2. &  37.  \\    
     &     &   & [S~{\sc ii}]	&  6731  &   10.  &  2. &  30.  \\    
     &     &   & He~{\sc i}	&  7065  &   2. &  1. &   7.  \\    
     &     &   & [Ar~{\sc iii}] &  7135  &   9. &  2. &  32.  \\    
     &     &   & [S~{\sc iii}]  &  9069  &   3.  &  1. &  27.  \\   
\hline
H-8&1.32e-15& 0.11       & [O~{\sc ii}]  &  3727  &  53.	&  2.  &  49.   \\
&$\pm$7e-17& $\pm$0.07&[Ne~{\sc iii}]& 3968  &   1.3	&  0.5  &   1.2   \\  
     &     &                &  H~{\sc i}     &  3889  &   4.7	&  0.5  &   4.4   \\
     &     &                & [Ne~{\sc iii}] &  3968  &   6.7	&  0.5  &   6.3   \\
     &     &		    & H$\delta$      &  4100  &  12.5	&  0.6  &  11.9   \\
     &     &		    & H$\gamma$      &  4340  &  32.	&  2.   &  31.   \\
     &     &		    & [O~{\sc iii}]  &  4363  &   1.7	&  0.3  &   1.7   \\
     &     &		    & [Fe~{\sc iii}] &  4702  &   6.3	&  0.5  &   6.3   \\
     &     &		    & H$\beta$       &  4861  & 100.	&  6.  & 100.   \\
     &     &		    & [O~{\sc iii}]  &  4959  &  19.	&  1.   &  19.   \\
     &     &		    & [O~{\sc iii}]  &  5007  &  55.	&  3.  &  56.   \\
     &     &                & [O~{\sc i}]    &  6300  &   1.7	&  0.5  &   1.9   \\	  
     &     &                & [N~{\sc ii}]   &  6548  &  11.	&  1.  &  12.   \\	
     &     &		    & H$\alpha$      &  6563  & 287.	& 15.   & 310.   \\
     &     &		    & [N~{\sc ii}]   &  6584  &  25.	&  2.  &  27.   \\
     &     &		    & [S~{\sc ii}]   &  6717  &  36. 	&  3.  &  39.   \\
     &     &		    & [S~{\sc ii}]   &  6731  &  26.		&  2.  &  28.   \\
     &     &                & [O~{\sc ii}]   &  7320  &   6.2	&  1.  &   6.9   \\
\hline
H-9&1.12e-15&1.1 &  H~{\sc i}  &    3686 & 11. & 0.5 &   5.1 \\
   &$\pm$6e-17&$\pm$0.1&  He~{\sc i}    & 3705 &  2.3 & 0.5 &   1.1 \\ 
     &     &	& [O~{\sc ii}]   & 3727  & 58.   &  2. &  28. \\
     &     &	&   He~{\sc i}  & 3785  & 20.   &  0.5&  10. \\   
     &     &	&   He~{\sc i}   & 3833  &  8.4 &  0.5&   4.4 \\		   
      &     &	&[Ne~{\sc iii}] &  3868  & 14. &  0.5 &   7.4\\  
     &     &    &  [S~{\sc ii}]  &  4068  & 12. &  0.5&   7.1   \\
     &     &    & H$\delta$	&  4100  & 21. &  1. &  13.   \\
     &     &    & H$\gamma$	&  4340  & 40. &  1. &  29.   \\
     &     &    & He~{\sc i}    &  4471  &  5.5 &  0.5&   4.3   \\
      &     &    & H$\beta$	&  4861  &100. &  4. & 100.   \\
     &     &    & [O~{\sc iii}] &  4959  &103. &  5.& 110.   \\
      &     &    & [O~{\sc iii}] &  5007  &320.  & 15. & 351.   \\
      &     &   & He~{\sc i}	&  5876  & 15. &  2. &  26. \\
     &     &   & [O~{\sc i}]    &  6300  & 30. &  4. &  59.   \\	 
       &     &   & H$\alpha$	&  6563  &287.  & 30. & 635.   \\   
     &     &   & [N~{\sc ii}]	&  6584  & 12. &  2. &  27.   \\   
      &     &   & He~{\sc i}	&  6678  &  7. &  1 &  15.  \\    
     &     &   & [S~{\sc ii}]	&  6717  & 11. &  2. &  26.  \\    
     &     &   & [S~{\sc ii}]	&  6731  &  8.2 &  1.&  19.  \\    
       &     &   & [Ar~{\sc iii}] &  7135  &  7.6 &  1.&  20.  \\ 
     &     &   & [O~{\sc ii}]   &  7320  &  1.4 &  0.5&   3.9  \\
     &     &   & [O~{\sc ii}]   &  7330  &  0.7 &  0.3&   1.9  \\  
     &     &   & [Ar~{\sc iii}] &  7751  &  1.7 &  0.3&   5.5  \\    
     &     &   & [S~{\sc iii}]  &  9069  &  4.6 &  2. &  24.4  \\    
\hline
H-10&3.5e-16& 0.16       & [O~{\sc ii}]  &  3727  &   46.   &  2.  &  42.   \\
&$\pm$2e-17&$\pm$0.05& H$\gamma$    &  4340  &   30.   &  1.  &  28.   \\  
     &     &		    & [O~{\sc iii}]  &  4363  &    0.4   &  0.2  &   0.4   \\
     &     &		    &  He~{\sc i}     &  4437  &    9.2   &  0.5  &	8.9   \\
      &     &		    & H$\beta$       &  4861  &  100.   &  4.  & 100.   \\
      &     &		    & [O~{\sc iii}]  &  4959  &   12.   &  0.6  &  12.   \\
     &     &		    & [O~{\sc iii}]  &  5007  &   27.    &  1. &  27.   \\
       &     &		    & H$\alpha$      &  6563  &  287.   & 15.  & 321.   \\
     &     &		    & [N~{\sc ii}]   &  6584  &   26.   &  2.  &  29.   \\
     &     &		    & [S~{\sc ii}]   &  6717  &   45.   &  3.  &  50.   \\
     &     &		    & [S~{\sc ii}]   &  6731  &   40.   &  3.  &  45.   \\
\hline
\end{tabular}
}
\end{minipage}
\label{tabPN_flux}
\end{table}

\begin{table}
\centering
\begin{minipage}{82mm}
{\tiny 
\contcaption{}
\begin{tabular}{@{}ccclclll@{}}
\hline
Id & F$_{{\rm H}\beta}$ & \cbeta\ & Ion & $\lambda$ (\AA) & I$_{\lambda}$ & $\Delta$F$_{\lambda}$ & F$_{\lambda}$ \\ 
   & $\Delta$F$_{\lambda}$  & $\Delta$\cbeta\ &     &		      & 	      &        (\%)		&  \\	       
\hline 
H-11&1.27e-14&0.0 & [O~{\sc ii}]&3727 & 24. &  1.&  24. \\
   &$\pm$6e-16&$\pm$0.1& [Ne~{\sc iii}] &  3868  &  4.3 &  0.2&   4.3\\ 
      &     &    &  H~{\sc i}	&  3889  &  3.8 &  0.2&   3.8   \\
     &     &    & [Ne~{\sc iii}] & 3968  &  7.0 &  0.3&   7.0   \\
       &     &    & H$\delta$	&  4100  & 11. &  0.5&  11.   \\
     &     &    & H$\gamma$	&  4340  & 28. &  1.&  28.   \\
     &     &    & He~{\sc i}    &  4471  &  1.3 &  0.3 &   1.3   \\
     &     &    & H$\beta$	&  4861  &100. &  3. & 100.   \\
     &     &    & He~{\sc i}	&  4922  & 10. &  0.5&  10.   \\
     &     &    & [O~{\sc iii}] &  4959  &122. &  6.& 122.   \\
     &     &    & [O~{\sc iii}] &  5007  &334. & 16.& 334.   \\
      &     &		  & He~{\sc i}     &  5876  &	3. &  1. &  3.  \\      
       &     &   & H$\alpha$	&  6563  & 223. & 14.&223.   \\   
     &     &   & [N~{\sc ii}]	&  6584  &   9.7 &  0.6&  9.7   \\   
    &     &   & He~{\sc i}	&  6678  &   3.3 &  0.2&  3.3  \\    
     &     &   & [S~{\sc ii}]	&  6717  &  15. &  1. & 15.  \\    
     &     &   & [S~{\sc ii}]	&  6731  &  11. &  0.7 & 11.  \\    
     &     &   & He~{\sc i}     &  7065  &   1.5 &  0.5 &  1.5   \\		      
     &     &   & [Ar~{\sc iii}] &  7135  &  10. &  1. & 10.  \\ 
      &     &   & [O~{\sc ii}]   &  7320  &   2.9 &  0.3 &  2.9  \\
     &     &   & [O~{\sc ii}]   &  7330  &   2.3 &  0.3 &  2.3  \\  
     &     &   & [Ar~{\sc iii}] &  7751  &   2.7 &  0.3 &  2.7  \\    
 \hline
H-12&2.0e-15&0.12  & [O~{\sc ii}]	   &  3727  &	27. &  1. &  25. \\
&$\pm$1e-16&$\pm$0.05 &He~{\sc i}   &  3833  &	 1.5 &  0.5 &   1.4  \\	       
     &     &		  & [Ne~{\sc iii}] &  3868  &	 7.1 &  0.3 &   6.7  \\      
     &     &		  & H~{\sc i}	   &  3889  &	 5.2 &  0.3 &   4.9  \\
     &     &		  & [Ne~{\sc iii}] &  3968  &	 7.9 &  0.4 &   7.4  \\      
     &     &		  & H$\delta$	   &  4100  &	13. &  0.6 &  13.  \\      
     &     &		  & H$\gamma$	   &  4340  &	32. &  2. &  31.   \\     
     &     &		  & [O~{\sc iii}]  &  4363  &	 2.9 &  0.2 &   2.8   \\     
     &     &		  & He~{\sc i}     &  4471  &	 3.0 &  0.2 &   2.9   \\     
      &     &		  & H$\beta$	   &  4861  &  100. &  5. & 100.  \\      
     &     &		  & [O~{\sc iii}]  &  4959  &  141. &  7. & 142.  \\      
     &     &		  & [O~{\sc iii}]  &  5007  &  330. & 16. & 333.  \\      
      &     &		  & He~{\sc i}     &  5876  &	15. &  1. &  16.  \\      
     &     &		  & [N~{\sc ii}]   &  6548  &	 3.8 &  0.3 &   4.2	\\    
     &     &		  & H$\alpha$	   &  6563  &  287. & 20. & 313.   \\     
     &     &		  & [N~{\sc ii}]   &  6584  &	 7.3 &  0.5 &   7.9   \\     
     &     &		  & He~{\sc i}     &  6678  &	 4.6 &  0.3 &   5.0	\\    
     &     &		  & [S~{\sc ii}]   &  6717  &	10. &  1. &  11.	\\    
     &     &		  & [S~{\sc ii}]   &  6731  &	10. &  1. &  11.	\\    
     &     &		  & He~{\sc i}     &  7065  &	 4.5 &  0.3 &   4.9   \\	
     &     &		  & [Ar~{\sc iii}] &  7135  &	 9.8 &  0.7 &  11.   \\	      
     &     &		  & [O~{\sc ii}]   &  7320  &	 4.3 &  0.3 &   4.8   \\	      
     &     &		  & [O~{\sc ii}]   &  7330  &	 3.8 &  0.3 &   4.3   \\	      
     &     &		  & [Ar~{\sc iii}] &  7751  &	 2.8 &  0.2 &   3.2   \\	      
\hline 
H-13&1.10e-15&0.37	  & [Ne~{\sc iii}] &  3968  &  18. &  1. &  15.\\
&$\pm$6e-17&$\pm$0.05  &  H$\delta$	   &  4100  &  12. &  0.5 &  10. \\
       &     &		  & H$\gamma$	   &  4340  &  34. &  2. &  30.  \\
     &     &		  & [O~{\sc iii}]  &  4363  &   2.6 &  0.2 &   2.3  \\
      &     &		  &He~{\sc i}	   &  4471  &   2.4 &  0.2 &   2.2   \\
     &     &		  & H$\beta$	   &  4861  &  100. &  5. & 100.   \\
     &     &		  & [O~{\sc iii}]  &  4959  & 109. &  5. & 111.   \\
     &     &		  & [O~{\sc iii}]  &  5007  & 338. & 15. & 350.   \\
     &     &		  & H$\alpha$	   &  6563  & 287. & 20. & 382.    \\
     &     &		  & [N~{\sc ii}]   &  6584  &   6.7 &  0.6 &   8.8   \\
     &     &		  & [S~{\sc ii}]   &  6717  &   4.4 &  0.4 &   5.9   \\
     &     &		  & [S~{\sc ii}]   &  6731  &   5.5 &  0.5 &   7.4   \\
\hline 
H-14&1.84e-15&0.38	& [O~{\sc ii}]   &  3727  &  57.  &  2. &  45.  \\ 
&$\pm$9e-17&$\pm$0.05& [Ne~{\sc iii}] &  3868  &  11.  &  0.5 &	8.9  \\
     &     &		 & H~{\sc i}	  &  3889  &   3.2  &  0.3 &	2.6  \\
     &     &		 & [Ne~{\sc iii}] &  3968  &  14.  &  0.6 &  12.  \\
     &     &		 & H$\delta$	  &  4100  &  11.  &  0.5 &	9.7  \\
     &     &		 & H$\gamma$	  &  4340  &  33.  &  2. &  29.  \\
     &     &		 & [O~{\sc iii}]  &  4363  &   3.1  &  0.2 &	2.8  \\
     &     &		 & H$\beta$	  &  4861  &  100.  &  6. & 100.   \\
     &     &		 & [O~{\sc iii}]  &  4959  & 125.  &  6. & 128.   \\
     &     &		 & [O~{\sc iii}]  &  5007  & 381.  & 15. & 394.   \\ 
     &     &		  & He~{\sc i}    &  5876  &  13.  &  1. &  16.  \\
     &     &		 & [N~{\sc ii}]   & 6548   &   6.7  &  0.6 &	8.7   \\
     &     &		 & H$\alpha$	  &  6563  & 287.  & 20. & 377.  \\
     &     &		 & [N~{\sc ii}]   &  6584  &   9.8  &  1. &  13.  \\
     &     &		 & [S~{\sc ii}]   &  6717  &  17.  &  1. &  23.  \\
     &     &		 & [S~{\sc ii}]   &  6731  &  14.  &  1. &  19.  \\  
 \hline
H-15&2.3e-15&0.0      &  He~{\sc i}    & 3587  &   3.2  &  0.3 &   3.2\\
&$\pm$1e-16&$\pm$0.01&He~{\sc i}  & 3613  &   6.1  &  0.3 &   6.1 \\
     &     &		 &	He~{\sc i}  & 3634  &   3.5  &  0.3 &   3.5  \\
     &     &		 &	H~{\sc i}  & 3659  &   4.1  &  0.3 &   4.1  \\
     &     &		 &	H~{\sc i}  & 3673  &   7.4  &  0.4 &   7.4  \\
     &     &		 & [O~{\sc ii}]   & 3727  &  34.   &  2. &  34. \\
     &     &		 &	H~{\sc i}  & 3750  &   5.3  &  0.3 &   5.3 \\
     &     &		 & [Ne~{\sc iii}] & 3868  &   3.5  &  0.3 &   3.5  \\
     &     &		 & H~{\sc i}	  & 3889  &   8.0  &  0.4 &   8.0  \\
     &     &		 & [Ne~{\sc iii}] & 3968  &  10.  &  0.5 &  10. \\
     &     &		& H$\delta$	  &  4100 &  13.  &  0.6 &  13.  \\
     &     &		 & H$\gamma$	  & 4340  &  34.  &  2. &  34.   \\
     &     &		 & [O~{\sc iii}]  &  4363 &   1.2  &  0.1 &   1.2  \\
     &     &		 & O~{\sc ii}	  & 4676  &   2.5  &  0.2 &   2.5  \\
     &     &		 & H$\beta$	  & 4861  & 100.  &  5. & 100.   \\
     &     &		 & [O~{\sc iii}]  & 4959  & 115.  &  6. & 115.   \\
     &     &		 & [O~{\sc iii}]  & 5007  & 339.  & 20.  & 339.   \\
     &     &		 & He~{\sc i}	  & 5876  &  11.  &  1. &  11.   \\
     &     &		 & [N~{\sc ii}]  & 6527  &   4.7  &  0.3 &   4.7  \\
     &     &		 & H$\alpha$	  & 6563  & 272.  & 15. & 272.  \\
      &     &		 & [S~{\sc ii}]   & 6717  &  12. &  1. &  12. \\
     &     &		 & [S~{\sc ii}]   & 6731  &   9.2  &  0.6 &   9.2  \\
\hline
\end{tabular}
}
\end{minipage}
\label{tabPN_flux}
\end{table}

\begin{table}
\centering
\begin{minipage}{82mm}
{\tiny 
\contcaption{}
\begin{tabular}{@{}ccclclll@{}}
\hline
Id & F$_{{\rm H}\beta}$ & \cbeta\ & Ion & $\lambda$ (\AA) & I$_{\lambda}$ & $\Delta$F$_{\lambda}$ & F$_{\lambda}$ \\ 
   &  $\Delta$F$_{\lambda}$ & $\Delta$\cbeta\ &     &                 &               &        (\%)   	     &  \\	     \hline
H-16&2.00e-15&0.26     & [O~{\sc ii}]   & 3727  &  43.  &   2. &  37. \\
&$\pm$2e-17&$\pm$0.05&	H~{\sc i}  &3750  &   3.4  &   0.5 &   2.9  \\
     &     &		 &	H~{\sc i}  & 3770  &   4.4  &   0.3 &   3.7 \\
     &     &		 &	He~{\sc i}  & 3785  &   2.6  &   0.3  &   2.2  \\
     &     &		 & H$\delta$	  &  4100 &  10.  &   0.5 &   9.2  \\
     &     &		 & H$\gamma$	  & 4340  &  33.  &   2. &  31.    \\
     &     &		 &  He~{\sc i}     & 4387  &   2.1  &   0.3 &   1.9  \\
     &     &		 & H$\beta$	  & 4861  & 100.  &   5. & 100.   \\
     &     &		 & [O~{\sc iii}]  & 4959  &  81.  &   4. &  82.   \\
     &     &		 & [O~{\sc iii}]  & 5007  & 259.  &  13. & 265.   \\
     &     &		 & [N~{\sc ii}]   & 6548  &   5.7  &   0.5 &   6.8  \\
     &     &		 & H$\alpha$   & 6563  & 287.  &  20.   & 346.  \\
     &     &		 &  [N~{\sc ii}]  & 6584  &  21.  &   2. &  25. \\
      &     &		 & [S~{\sc ii}]   & 6717  &  21.  &   2. &  25.  \\
     &     &		 & [S~{\sc ii}]   & 6731  &  18.  &   2. &  22.  \\
\hline 
H-17&4.3e-15&0.71  &    [O~{\sc ii}]     &  3727  &  68. &  2. &  43.  \\ 	  
&$\pm$2e-16&$\pm$0.05 &     He~{\sc i}    &  3833  &	5.1 &  0.3 &   3.3  \\
      &     &		  & [Ne~{\sc iii}] &  3868  &	4.8 &  0.3 &   3.2  \\      
     &     &		  & H~{\sc i}	   &  3889  &	4.3 &  0.3 &   2.9  \\
     &     &		  & [Ne~{\sc iii}] &  3968  &  11. &  0.4 &   8.  \\      
     &     &		  & H$\gamma$	   &  4340  &  45. &  2. &  36.   \\     
     &     &		  & [O~{\sc iii}]  &  4363  &	0.9 &  0.3 &   0.7   \\     
     &     &		  & He~{\sc i}     &  4471  &	3.7 &  0.3 &   3.2   \\     
     &     &		  & H$\beta$	   &  4861  & 100. &  5. & 100.  \\      
     &     &		  & [O~{\sc iii}]  &  4959  &  73. &  4. &  76.  \\      
     &     &		  & [O~{\sc iii}]  &  5007  & 214. & 10.& 227.  \\      
     &     &		  & He~{\sc i}     &  5876  &	9.4 &  0.7 &  13.  \\      
     &     &		  & [N~{\sc ii}]   &  6548  &	5.8 &  0.6 &   9.5    \\    
     &     &		  & H$\alpha$	   &  6563  & 287. & 30. & 478.   \\     
     &     &		  & [N~{\sc ii}]   &  6584  &  13. &  1. &  21.   \\     
     &     &		  & [S~{\sc ii}]   &  6717  &  10. &  1. &  17.    \\    
     &     &		  & [S~{\sc ii}]   &  6731  &	7.3 &  1. &  12.    \\    
      &     &		  & [Ar~{\sc iii}] &  7135  &	5.9 &  1. &  11.   \\	     
\hline
H-18&2.6e-16&0.0     & H$\delta$	   &  4100 &   11. &  1. &  11.  \\
&$\pm$5e-17&$\pm$0.1 & H$\gamma$ &  4340  &  26. &  2. &  26.   \\	      
     &     &		  & [O~{\sc iii}]  &  4363  &   8.0 &  1. &   8.0   \\     
     &     &		  & H$\beta$	   &  4861  & 100. &  5. & 100.  \\      
     &     &		  & [O~{\sc iii}]  &  4959  & 279. & 14. & 279.  \\      
     &     &		  & [O~{\sc iii}]  &  5007  & 784. & 40. & 784.  \\      
     &     &		  & He~{\sc i}     &  5876  &  13. &  3. &  13.  \\      
     &     &		  & H$\alpha$	 &  6553  & 245. & 15. & 245.    \\    
     &     &		  & [N~{\sc ii}]     &  6584  &  22.8 &  1.46 &  22.8   \\     
\hline
H-19&7.7e-15&0.0     &[O~{\sc ii}]     &  3727  &   9.2 &  0.5 &	9.2  \\
&$\pm$4e-16&$\pm$0.1   & [Ne~{\sc iii}] &  3868  &   8.3 &  0.5 &	8.3  \\      
      &     &		  & [Ne~{\sc iii}] &  3968  &   9.4 &  0.5 &	9.4  \\  
     &     &              & H$\delta$	   &  4100 &   12. &  0.6 &  12.  \\    
     &     &		  & H$\gamma$	   &  4340  &  31. &  2. &  31.   \\     
     &     &		  & [O~{\sc iii}]  &  4363  &   4.2 &  0.2 &	4.2   \\     
     &     &		  & He~{\sc i}     &  4471  &   2.9 &  0.2 &	2.9   \\     
      &     &		  & H$\beta$	   &  4861  & 100. &  5. & 100. \\      
     &     &		  & [O~{\sc iii}]  &  4959  & 163. &  8. & 163.  \\      
     &     &		  & [O~{\sc iii}]  &  5007  & 396. & 20. & 396.  \\      
     &     &		  & He~{\sc i}     &  5876  &  10. &  1. &  10.  \\      
     &     &		  & H$\alpha$	   &  6563  & 270. & 15. & 270.   \\     
     &     &		  & [N~{\sc ii}]   &  6584  &   2.7 &  1. &	2.7   \\     
     &     &		  & [S~{\sc ii}]   &  6678  &   3.6 &  1. &	3.6    \\    
     &     &		  & [S~{\sc ii}]   &  6717  &   3.1 &  1. &	3.1    \\    
     &     &		  & He~{\sc i}     &  6731  &   2.6 &  1. &	2.6	\\    
     &     &		  &  He~{\sc i}     &  7065  &   2.2 &  1. &	2.2   \\       
     &     &		  & [Ar~{\sc iii}] &  7135  &   6.9 &  2. &	6.9   \\	     
\hline
H-20&9.0e-15&0.50    &[O~{\sc ii}]     &  3727  &  29. &  1. &  21.  \\
&$\pm$4e-16&$\pm$0.05  &  He~{\sc i}     &  3833  &   2.5 &  0.5 &   1.8  \\       
     &     &		  & [Ne~{\sc iii}] &  3868  &   8.4 &  0.3 &   6.3  \\  
     &     &		  & H~{\sc i}	   &  3889  &   5.1 &  0.2 &   3.9  \\ 	
     &     &		  & [Ne~{\sc iii}] &  3968  &  11. &  0.4 &   8.2  \\  
     &     &              & H$\delta$	   &  4100 &   13. &  0.5 &  11.  \\    
     &     &		  & H$\gamma$	   &  4340  &  37. &  2. &  32.  \\     
     &     &		  & [O~{\sc iii}]  &  4363  &   3.7 &  0.5 &   3.2  \\     
     &     &		  & He~{\sc i}     &  4471  &   4.7 &  0.3 &   4.2  \\     
     &     &		  & H$\beta$	   &  4861  & 100. &  5. & 100.  \\      
     &     &		  & [O~{\sc iii}]  &  4959  & 158. &  8. & 163.  \\      
     &     &		  & [O~{\sc iii}]  &  5007  & 257. & 13. & 268.  \\      
     &     &		  & He~{\sc i}     &  5876  &  12. &  1. &  15.  \\      
     &     &		  & [O~{\sc i}]    &  6300  &   1.9 &  0.3 &   2.6  \\     
     &     &		  & [S~{\sc iii}]  &  6312  &   2.0 &  0.3 &   2.7  \\    
      &     &		  & H$\alpha$	   &  6563  & 287. & 25. & 412.  \\     
     &     &		  & [N~{\sc ii}]   &  6584  &  12. &  1. &  17.  \\     
     &     &		  & He~{\sc i}     &  6678  &   5.0 &  0.5 &   7.8  \\    
     &     &		  & [S~{\sc ii}]   &  6717  &  12. &  1. &  17.  \\    
     &     &		  & [S~{\sc ii}]   &  6731  &   9.3 &  1. &  13.  \\    
     &     &		  &  He~{\sc i}    &  7065  &   3.5 &  0.5 &   5.4  \\       
      &     &		  & [Ar~{\sc iii}] &  7135  &  10. &  1. &  16. \\ 	    
     &     &              & [O~{\sc ii}]   &  7320  &   5.8 &  1. &   9.2  \\
     &     &              & [O~{\sc ii}]   &  7330  &   3.4 &  1. &   5.5  \\  
     &     &              & [S~{\sc iii}]  &  9069  &  22. &  3. &  46.  \\  
\hline
H-21&1.8e-15&0.0     & [O~{\sc ii}]   & 3727  &   40.  &  2. &  40.  \\
&$\pm$1e-15&$\pm$0.05 & H$\gamma$ & 4340  &   30.  &  2. &  30.  \\
     &     &		 &  [O~{\sc iii}] & 4363  &    1.0  &  0.5 &   1.0  \\
     &     &		 & H$\beta$	  & 4861  &  100.  &  5. & 100.  \\
     &     &		 & [O~{\sc iii}]  & 4959  &   52.  &  3. &  52.  \\
     &     &		 & [O~{\sc iii}]  & 5007  &  138.  &  7. & 138.  \\
     &     &		 & He~{\sc i}     & 5876  &    9.3  &  0.6 &   9.3  \\      
     &     &		 & [N~{\sc ii}]   & 6548  &    4.0  &  0.5 &   4.0  \\
     &     &		 & H$\alpha$	  & 6563  &  275.  & 15. & 275.  \\
     &     &		 &  [N~{\sc ii}]  & 6584  &   18.  &  2. &  18.  \\
     &     &		 & He~{\sc i}     & 6678  &    4.7  &  0.3 &   4.7  \\    
     &     &		 & [S~{\sc ii}]   & 6717  &   30.  &  2. &  30.  \\
     &     &		 & [S~{\sc ii}]   & 6731  &   24.  &  2. &  24.  \\
     &     &             & [O~{\sc ii}]   & 7330  &   10.  &  1. &  10.  \\  
     &     &             & He~{\sc i}     & 7065  &    9.8  &  0.6 &   9.8  \\ 	      
\hline
\end{tabular}
}
\end{minipage}
\label{tabPN_flux}
\end{table}
\begin{table}
\centering
\begin{minipage}{82mm}
{\tiny 
\contcaption{}
\begin{tabular}{@{}ccclclll@{}}
\hline
Id & F$_{{\rm H}\beta}$ & \cbeta\ & Ion & $\lambda$ (\AA) & I$_{\lambda}$ & $\Delta$F$_{\lambda}$ & F$_{\lambda}$ \\ 
   &  $\Delta$F$_{\lambda}$ & $\Delta$\cbeta\ &     &                 &               &        (\%)   	     &  \\	
\hline
H-22&8.3e-15&0.14     &[O~{\sc ii}]     &  3727  &  26. &  2. &  24.  \\
&$\pm$4.e-16&$\pm$0.05  &   He~{\sc i}      &  3833  &   1.7 &  0.5 &   1.5  \\       
     &     &		  & [Ne~{\sc iii}] &  3868  &   4.8 &  0.5 &   4.4  \\  
     &     &		  & H~{\sc i}	   &  3889  &   4.5 &  0.5 &   4.2  \\ 	
     &     &		  & [Ne~{\sc iii}] &  3968  &   6.7 &  0.5 &   6.2  \\  
     &     &              & H$\delta$	   &  4100  &  12. &  1. &  11.  \\    
      &     &		  & H$\gamma$	   &  4340  &  33. &  2. &  31.  \\     
     &     &		  & [O~{\sc iii}]  &  4363  &   2.3 &  0.3 &   2.2  \\     
     &     &		  & He~{\sc i}     &  4471  &   3.9 &  0.5 &   3.8  \\     
     &     &		  & H$\beta$	   &  4861  & 100. &  5. & 100.  \\      
     &     &		  & [O~{\sc iii}]  &  4959  & 126. &  6. & 127.  \\      
     &     &		  & [O~{\sc iii}]  &  5007  & 369. & 20. & 374.  \\      
      &     &	 	  & [N~{\sc ii}]   &  6548  &   3.8 &  1. &   4.2  \\
     &     &		  & H$\alpha$	   &  6563  & 287. & 20. & 318.  \\     
     &     &		  & [N~{\sc ii}]   &  6584  &  13. &  1. &  15.  \\     
     &     &		  & He~{\sc i}     &  6678  &   3.9 &  0.5 &   4.4  \\    
     &     &		  & [S~{\sc ii}]   &  6717  &  11. &  1. &  12.  \\    
     &     &		  & [S~{\sc ii}]   &  6731  &   8.2 &  0.6 &   9.2  \\    
\hline
H-23&2.468e-14&0.19     &[O~{\sc ii}]     &  3727  &  13. &  0.6 &  11.  \\
&$\pm$1.2e-15&$\pm$0.05  & He~{\sc i}   &  3833  &   1.2 &  0.5 &   1.1  \\
      &     &		  & [Ne~{\sc iii}] &  3868  &   5.7 &  0.5 &   5.1  \\      
     &     &		  & H~{\sc i}	   &  3889  &   4.5 &  0.5 &   4.0  \\ 	
     &     &		  & [Ne~{\sc iii}] &  3968  &   7.7 &  0.5 &   7.0  \\  
     &     &              & H$\delta$	   &  4100 &   11. &  0.5 &  10.  \\    
     &     &		  & [O~{\sc iii}]  &  4363  &   1.5 &  0.3 &   1.4  \\     
     &     &		  & He~{\sc i}     &  4471  &   3.7 &  0.4 &   3.6  \\     
     &     &		  & H$\beta$	   &  4861  & 100. &  5. & 100.  \\      
     &     &		  & [O~{\sc iii}]  &  4959  & 108. &  5. & 109.  \\      
     &     &		  & [O~{\sc iii}]  &  5007  & 322. & 15. & 328.  \\      
     &     &		  & He~{\sc i}     &  5876  &  12. &  2. &  13.  \\      
     &     &		  & H$\alpha$	   &  6563  & 287. & 29. & 330.  \\     
     &     &		  & [N~{\sc ii}]   &  6584  &   7.1 &  0.5 &   8.1  \\     
     &     &		  & He~{\sc i}     &  6678  &   5.2 &  0.4 &   6.0  \\    
     &     &		  & [S~{\sc ii}]   &  6717  &   5.6 &  0.4 &   6.5  \\    
     &     &		  & [S~{\sc ii}]   &  6731  &   4.4 &  0.3 &   5.1  \\    
     &     &		  &  He~{\sc i}    &  7065  &   2.7 &  0.2 &   3.1  \\       
     &     &		  & [Ar~{\sc iii}] &  7135  &  12. &  1. &  15.  \\ 	    
     &     &		  & [Ar~{\sc iii}] &  7751  &   2.9 &  0.5 &   3.5  \\       
     &     &		  &  [S~{\sc iii}]   &  9530  &  20. &  2. &  28.  \\       
\hline
H-24&9.5e-16&0.32     &[O~{\sc ii}]     &  3727  &   18.  &  1. &  15.  \\
&$\pm$5.e-17&$\pm$0.05 &  H~{\sc i}&  3797  &    3.1  &  0.5 &   2.5  \\	
      &     &		  & [Ne~{\sc iii}] &  3968  &    5.0  &  1. &   4.2  \\  
     &     &              & H$\delta$	   &  4100  &   14.  &  1. &  12.  \\    
     &     &              & H$\gamma$ 	   &  4340  &   32.  &  2. &  30.  \\    
     &     &		  & He~{\sc i}     &  4471  &    1.1  &  0.5 &   1.0  \\     
      &     &		  & H$\beta$	   &  4861  &  100.  &  5. & 100.  \\      
     &     &		  & [O~{\sc iii}]  &  4959  &   71.  &  4. &  72.  \\      
     &     &		  & [O~{\sc iii}]  &  5007  &  220.  & 10. & 226.  \\      
      &     &		  & He~{\sc i}     &  5876  &    6.2  &  0.5 &   7.2  \\      
     &     &	 	  & [N~{\sc ii}]   &  6548  &    8.5  &  1. &  11.  \\
     &     &		  & H$\alpha$	   &  6563  &  287.  & 20. & 360.  \\     
     &     &		  & [N~{\sc ii}]   &  6584  &   26.  &  2. &  33.  \\     
     &     &		  & [S~{\sc ii}]   &  6717  &   27.  &  2. &  34.  \\    
     &     &		  & [S~{\sc ii}]   &  6731  &   21.  &  2. &  27.  \\    
     &     &		  & [Ar~{\sc iii}] &  7135  &    8.6  &  2. &  11.  \\	      
\hline
H-25&9.0e-15&0.0     &[O~{\sc ii}]    &  3727  &   9.0 &  0.5 &	9.0  \\
&$\pm$5.e-16&$\pm$0.1  & [Ne~{\sc iii}] &  3968  &   6.5 &  0.3 &	6.5  \\  
      &     &		  & H~{\sc i}	   &  3889  &   3.7 &  0.3 &	3.7  \\ 	
     &     &		  & [Ne~{\sc iii}] &  3968  &   7.7 &  0.4 &	7.7  \\  
     &     &              & H$\delta$	   &  4100  &   8.7 &  0.4 &	8.7  \\    
     &     &		  & H$\gamma$	   &  4340  &  18. &  3. &  18.  \\     
     &     &		  & [O~{\sc iii}]  &  4363  &   3.1 &  0.5 &	3.1  \\     
     &     &		  & H$\beta$	   &  4861  & 100. &  8. & 100.  \\      
     &     &		  & [O~{\sc iii}]  &  4959  & 175. &  9. & 175.  \\      
     &     &		  & [O~{\sc iii}]  &  5007  & 588. & 30. & 588.  \\      
     &     &		  & He~{\sc i}     &  5876  &  16. &  2. &  16.  \\      
     &     &		  & [O~{\sc i}]    &  6300  &   1.6 &  1.&	1.6  \\     
     &     &		  & [S~{\sc iii}]  &  6312  &   2.3 &  1. &	2.3  \\    
     &     &		  & H$\alpha$	   &  6563  & 276. & 15. & 276.  \\     
     &     &		  & [N~{\sc ii}]   &  6584  &   8.4 &  0.5 &	8.4  \\     
     &     &		  & He~{\sc i}     &  6678  &   4.6 &  0.5 &	4.6  \\    
     &     &		  & [S~{\sc ii}]   &  6717  &  13. &  2. &  13.  \\    
     &     &		  & [S~{\sc ii}]   &  6731  &  10. &  1. &  10.  \\    
     &     &		  &  He~{\sc i}    &  7065  &   5.1 &  0.5 &	5.1  \\       
     &     &		  & [Ar~{\sc iii}] &  7135  &  12. &  2. &  12.  \\ 	    
\hline
\hline
\end{tabular}
}
\end{minipage}
\label{tabPN_flux}
\end{table}

\label{lastpage}


\begin{thebibliography}{}
\bibitem[Annibali et al.(2015)]{Annibali15} Annibali, F., Tosi, M., Pasquali, A., et al.\ 2015, AJ, 150, 143 

\bibitem[Arellano-C{\'o}rdova et al.(2015)]{ac15} 
Arellano-C{\'o}rdova, K.~Z., Rodr{\'{\i}}guez, M., Mayya, Y.~D., 
\& Rosa-Gonz{\'a}lez, D.\ 2015, arXiv:1510.07757 

\bibitem[Belczy{\'n}ski et al.(2000)]{bel00} Belczy{\'n}ski, K., Miko{\l}ajewska, J., Munari, U., Ivison, R.~J., \& Friedjung, M.\ 2000, A\&AS, 146, 407 


\bibitem[\protect\citeauthoryear{Benjamin et al.}{1999}]{benjamin99} 
Benjamin R. A., Skillman E. D., Smits D. P., 1999, ApJ, 514, 307

\bibitem[\protect\citeauthoryear{B\"ohringer et al.}{2000}]{bohringer00}
B\"ohringer H., Voges W., Huchra J.P., McLean B., Giacconi R., Rosati P., Burg R., Mader J., Schuecker P., Simi{\c c} D., Komossa S., Reiprich T.H., Retzlaff J., Tr{\"u}mper J., 2000, ApJS, 129, 435

\bibitem[Bresolin et al.(2009)]{Bresolin09} Bresolin, F., Gieren, 
W., Kudritzki, R.-P., et al.\ 2009, ApJ, 700, 309 

\bibitem[Carlos Reyes et al.(2015)]{CR15} Carlos Reyes, R.~E., Reyes Navarro, F.~A., Mel{\'e}ndez, J., Steiner, J., \& Elizalde, F.\ 2015, RevMex, 51, 135 

\bibitem[\protect\citeauthoryear{Carranza \& Ageeuro}{1988}]{CA88}
Carranza G. I., Agueero E. L., 1989, Ap\&SS, 152, 279

\bibitem[\protect\citeauthoryear{Castro et al.}{2008}]{castro08}
Castro N., Herrero A., Garcia M., Trundle C., Bresolin F., Gieren W.,
 Pietrzy\'nski G., Kudritzki R.-P., Demarco R.,  2008, A\&A, 
 485, 41
 
 \bibitem[Clayton et al.(2015)]{clayton15} Clayton, G.~C., Gordon, K.~D., Bianchi, L.~C., et al.\ 2015, ApJ, 815, 14 



\bibitem[\protect\citeauthoryear{Clegg}{1987}]{clegg87}
Clegg R. E. S., 1987, MNRAS, 229, 31

\bibitem[Cote et al.(1997)]{cote97} Cote, S., Freeman, K.~C., Carignan, C., \& Quinn, P.~J.\ 1997, AJ, 114, 1313 


\bibitem[Croxall et al.(2009)]{Croxall09} Croxall, K.~V., van Zee, L., Lee, H., et al.\ 2009, ApJ, 705, 723-738 

\bibitem[Dale et al.(2009)]{Dale09} Dale, D.~A., Cohen, S.~A., Johnson, L.~C., et al.\ 2009, ApJ, 703, 517 

\bibitem[Davidge(2005)]{Davidge05} Davidge, T.~J.\ 2005, ApJ, 622, 279 

\bibitem[Dettmar \& Heithausen(1989)]{DH89} Dettmar, R.-J., \& Heithausen, A.\ 1989, ApJL, 344, L61 

\bibitem[\protect\citeauthoryear{de Vaucouleurs}{1961}]{dV61} 
de Vaucouleurs G., 1961, ApJ, 133,405

\bibitem[\protect\citeauthoryear{de Vaucouleurs et al.}{1968}]{dV68}
de Vaucouleurs G., de Vaucouleurs A., \& Freeman K. C., 1968,
MNRAS, 139,425

\bibitem[Edmunds \& Pagel(1978)]{EP78} Edmunds, M.~G., \& Pagel, B.~E.~J.\ 1978, MNRAS, 185, 77P 

\bibitem[\protect\citeauthoryear{Engelbracht et al.}{2004}]{eng04}
Engelbracht C.W., Gordon K.D., Bendo G.J., P{\'e}rez-Gonz{\'a}lez P.G., Misselt K.A., Rieke G.H., Young E.T., Hines D.C., Kelly D.M., Stansberry J.A., Papovich C., Morrison J.E., Egami E., Su K.Y.L., Muzerolle J., Dole H., Alonso-Herrero A., Hinz J.L., Smith P.S., Latter W.B., Noriega-Crespo A., Padgett D.L., Rho J., Frayer D.T., Wachter S., 2004, ApJS, 154, 248

\bibitem[\protect\citeauthoryear{Ferguson, Wyse, \& Gallagher}{1996}]{ferguson96}
Ferguson A. M. N., Wyse R. F. G., Gallagher J. S., 1996, AJ, 112, 2567

\bibitem[Gazak et al.(2015)]{gazak15} Gazak, J.~Z., Kudritzki, R., Evans, C., et al.\ 2015, ApJ, 805, 182 


\bibitem[Gon{\c c}alves et al.(2007)]{goncalves07} Gon{\c c}alves, D.~R., Magrini, L., Leisy, P., \& Corradi, R.~L.~M.\ 2007, MNRAS, 375, 715 
\bibitem[Gon{\c c}alves et al.(2012)]{goncalves12} Gon{\c c}alves, D.~R., Magrini, L., Martins, L.~P., Teodorescu, A.~M., \& Quireza, C.\ 2012, MNRAS, 419, 854 
\bibitem[Gon{\c c}alves et al.(2014)]{goncalves14} Gon{\c c}alves, D.~R., Magrini, L., Teodorescu, A.~M., \& Carneiro, C.~M.\ 2014, MNRAS, 444, 1705 

\bibitem[Grevesse et al.(2007)]{grevesse07} Grevesse, N., Asplund, M., \& Sauval, A.~J.\ 2007, Space Science Reviews, 130, 105 

\bibitem[Haurberg et al.(2013)]{Haurberg13} Haurberg, N.~C., Rosenberg, J., \& Salzer, J.~J.\ 2013, ApJ, 765, 66 

\bibitem[\protect\citeauthoryear{Heithausen \&  Dettmar}{1990}]{HD90}
Heithausen A., Dettmar R.-J.,  1990, NASCP, 3084,68

\bibitem[\protect\citeauthoryear{Hoopes, Walterbos, \& Greenawalt}{1996}]{hoopes96}
Hoopes C. G., Walterbos R. A. M., Greenwalt B. E.,  1996, AJ, 112, 1429

\bibitem[\protect\citeauthoryear{Hamuy et al.}{1992}]{hamuy92}
Hamuy M., Walker A. R., Suntzeff N. B., Gigoux P., Heathcote S. R., 
Phillips M. M.,  1992, PASP, 104, 533

\bibitem[Hosek et al.(2014)]{hosek14} Hosek, M.~W., Jr., Kudritzki, R.-P., Bresolin, F., et al.\ 2014, ApJ, 785, 151 

	  
\bibitem[\protect\citeauthoryear{Hummel, Dettmar, \& Wielebinski}{1986}]{hummel86}
Hummel E., Dettmar R.-J., Wielebinski R., 1986, A\&A, 166,97

\bibitem[\protect\citeauthoryear{Hunter et al.}{2007}]{hunter07}
Hunter I., Dufton P. L., Smartt S. J., Ryans R. S. I., Evans C. J.,
 Lennon D. J., Trundle C., Hubeny I., Lanz T., 2007. A\&A, 466,277
 
 
\bibitem[\protect\citeauthoryear{Izotov et al.}{2006}]{izotov06}  	
Izotov Y. I., Stasi\'nska G., Meynet G., Guseva N. G., Thuan T. X., 
2006, A\&A, 448, 955

\bibitem[\protect\citeauthoryear{Kiszkurno-Koziej}{1988}]{kis88}
Kiszkurno-Koziej E.,  1988, A\&A, 196, 26

\bibitem[Kobulnicky \& Skillman(1997)]{KS97} Kobulnicky, H.~A., \& Skillman, E.~D.\ 1997, ApJ, 489, 636 

\bibitem[Kudritzki et al.(2008)]{k08} Kudritzki, R.-P., Urbaneja, M.~A., Bresolin, F., et al.\ 2008, ApJ, 681, 269-289 


\bibitem[Kudritzki et al.(2016)]{k16} Kudritzki, R., Urbaneja, M., Castro, N., et al.\ 2016, arXiv:1607.04325, K16


\bibitem[Jarrett et al.(2003)]{Jarrett03} Jarrett, T.~H., Chester, T., Cutri, R., Schneider, S.~E., \& Huchra, J.~P.\ 2003, AJ, 125, 525 


\bibitem[Jerjen et al.(2000)]{jerjen00} Jerjen, H., Binggeli, B., \& Freeman, K.~C.\ 2000, AJ, 119, 593 



\bibitem[Izotov et al.(1997)]{IT97} Izotov, Y.~I., Thuan, T.~X., \& Lipovetsky, V.~A.\ 1997, ApJS, 108, 1 

\bibitem[Lagos \& Papaderos(2013)]{LP13} Lagos, P., \& Papaderos, P.\ 2013, Advances in Astronomy, 2013, 1 

\bibitem[Leaman et al.(2014)]{Leaman14} Leaman, R., Venn, K., Brooks, A., et al.\ 2014, MemSAIt, 85, 504 


\bibitem[Lee et al.(2006)]{Lee06} Lee, H., Skillman, E.~D., \& Venn, K.~A.\ 2006, ApJ, 642, 813 


\bibitem[\protect\citeauthoryear{Liedahl,Osterheld \& Goldstein}{1995}]{liedahl95}
Liedahl D.A., Osterheld A.L., Goldstein W.H., 1995, ApJL, 438, 115

\bibitem[Magrini \& Gon{\c c}alves(2009)]{MG09} Magrini, L., \& Gon{\c c}alves, D.~R.\ 2009, MNRAS, 398, 280 

\bibitem[Magrini et al.(2005)]{magrini05} Magrini, L., Leisy, P., Corradi, R.~L.~M., et al.\ 2005, A\&A, 443, 115 

\bibitem[Magrini et al.(2007)]{Magrini07} Magrini, L., V{\'{\i}}lchez, J.~M., Mampaso, A., Corradi, R.~L.~M., \& Leisy, P.\ 2007, A\&A, 470, 865 

\bibitem[Magrini et al.(2009)]{magrini09} Magrini, L., Stanghellini, L., \& Villaver, E.\ 2009, ApJ, 696, 729 

\bibitem[Marino et al.(2013)]{marino13} Marino, R.~A., Rosales-Ortega, F.~F., S{\'a}nchez, S.~F., et al.\ 2013, A\&A, 559, A114 

\bibitem[\protect\citeauthoryear{Mathis}{1990}]{mathis90}
Mathis J. S., 1990, ARA\&A, 28, 37

\bibitem[\protect\citeauthoryear{Oshima}{2002}]{2002Oshima} Oshima, T., Mitsuda, K., Ota, N., \& Yamasaki, N.\ 2002, in IKEuchi S., Hearnshaw J., Hanawa T., eds, Proc. IAU 8th Asian-Pacific Regional Meeting, Volume II, Vol. 2. Pedagogical Univ. Press, p. 287

\bibitem[\protect\citeauthoryear{Osterbrock \& Ferland}{2006}]{osterbrock06}
Osterbrock D. E. \& Ferland G. J., in Astrophysics of gaseous nebulae and active galactic nuclei, 2nd. ed.
 Sausalito, CA: University Science Books, 2006

\bibitem[\protect\citeauthoryear{Otte \& Dettmar}{1999}]{OD99}
Otte B., Dettmar R.-J.,  1999, A\&A, 343, 705

\bibitem[Patrick et al.(2015)]{patrik15} Patrick, L.~R., Evans, C.~J., Davies, B., et al.\ 2015, ApJ, 803, 14 

\bibitem[Peimbert \& Peimbert(2010)]{pp10} Peimbert, A., \& Peimbert, M.\ 2010, ApJ, 724, 791 

\bibitem[Peimbert \& Torres-Peimbert(1974)]{ptp74} Peimbert, M., \& Torres-Peimbert, S.\ 1974, ApJ, 193, 327 

\bibitem[\protect\citeauthoryear{Persic \& Rephaeli}{2007}]{PR07}
Persic M., Rephaeli Y., 2007, A\&A, 463, 481

\bibitem[\protect\citeauthoryear{Pietrzy\'nski et al.}{2006}]{piet06}
Pietrzy\'nski G., Gieren W., Soszyński I., Udalski A., Bresolin F., et al.,
 2006, AJ, 132, 2556

\bibitem[Pilyugin et al.(2003)]{Pilyugin03} Pilyugin, L.~S., Thuan, T.~X., \& V{\'{\i}}lchez, J.~M.\ 2003, A\&A, 397, 487 

\bibitem[Pilyugin et al.(2014)]{Pilyugin14} Pilyugin, L.~S., Grebel, E.~K., Zinchenko, I.~A., \& Kniazev, A.~Y.\ 2014, AJ, 148, 134 

\bibitem[Pilyugin et al.(2015)]{Pilyugin15} Pilyugin, L.~S., Grebel, E.~K., \& Zinchenko, I.~A.\ 2015, MNRAS, 450, 3254 

\bibitem[\protect\citeauthoryear{Puche, Carignan \& Wainscoat}{1991}]{puche91}
Puche D., Carignan C., Wainscoat R. J.,  1991, AJ, 101,447


\bibitem[Renzini \& Voli(1981)]{RV81} Renzini, A., \& Voli, M.\ 1981, A\&A, 94, 175 


\bibitem[Sandage \& Tammann(1987)]{ST87} 
Sandage, A., \& Tammann, G.~A.\ 1987, Carnegie Institution of Washington Publication, Washington: Carnegie Institution, 1987, 2nd ed.,  

\bibitem[Schmid(1989)]{schmid89} Schmid, H.~M.\ 1989, A\&A, 211, L31 



\bibitem[Skillman(1998)]{skillman98} Skillman, E.~D.\ 1998, Stellar astrophysics for the local group: VIII Canary Islands Winter School of Astrophysics, 457 

\bibitem[Stanghellini et al.(2015)]{stanghellini15} Stanghellini, L., Magrini, L., \& Casasola, V.\ 2015, ApJ, 812, 39 

\bibitem[Stasi{\'n}ska(1982)]{stasinska82} Stasi{\'n}ska, G.\ 1982, AAPS, 48, 299 


\bibitem[Stasi{\'n}ska et al.(2013)]{Stasinska13} Stasi{\'n}ska,
G., Pe{\~n}a, M., Bresolin, F., \& Tsamis, Y.~G.\ 2013, A\&A, 552,
A12

\bibitem[\protect\citeauthoryear{Stobbart, Roberts \& Warwick}{2006}]{stobbart06}
Stobbart A. M., Roberts T. P., and Warwick R. S., 2006, MNRAS, 370, 25

\bibitem[\protect\citeauthoryear{Strickland}{2004}]{strickland04}
Strickland D.K., 2004, IAUS, 222, 249

\bibitem[Tanaka et al.(2011)]{tanaka11} Tanaka, M., Chiba, M., Komiyama, Y., Guhathakurta, P., \& Kalirai, J.~S.\ 2011, ApJ, 738, 150 

\bibitem[Tikhonov et al.(2005)]{tikhonov05} Tikhonov, N.~A., Galazutdinova, O.~A., \& Drozdovsky, I.~O.\ 2005, A\&A, 431, 127 


\bibitem[\protect\citeauthoryear{T\"ullmann et al.}{2003}]{tullmann03}
T\"ullmann R., Rosa M. R., Elwert T., Bomans D. J., Ferguson A. M. N., Dettmar R.-J., 2003, A\&A, 412, 69

\bibitem[\protect\citeauthoryear{T\"ullmann \& Rosa}{2004}]{TR04} 
T\"ullmann R., Rosa M. R., 2004, A\&A, 416, 243


\bibitem[van Zee et al.(1998)]{vZ98} van Zee, L., Salzer, J.~J., \& Haynes, M.~P.\ 1998, ApJL, 497, L1 

\bibitem[Venn et al.(2004)]{venn04} Venn, K.~A., Irwin, M., Shetrone, M.~D., et al.\ 2004, AJ, 128, 1177 

\bibitem[Vincenzo et al.(2016)]{vincenzo16} Vincenzo, F., Belfiore, F., Maiolino, R., Matteucci, F., \& Ventura, P.\ 2016, MNRAS, 458, 3466 


\bibitem[\protect\citeauthoryear{Webster \& Smith}{1983}]{WS83}
Webster B. L., Smith M. G., 1983, MNRAS, 204, 743

\bibitem[\protect\citeauthoryear{Westmeier et al.}{2013}]{westmeier13}
Westmeier T., Koribalski B. S., Braun R., 2013, MNRAS, 434, 3511

\bibitem[\protect\citeauthoryear{Whiting}{1999}]{whiting99}
Whiting A. B., 1999, AJ, 117, 202

\bibitem[\protect\citeauthoryear{Willingale}{2003}]{2003Willingale} Willingale R., Hands A.~D.~P., Warwick R.~S., Snowden S.~L., Burrows D.~N., 2003, MNRAS, 343, 995
 
\bibitem[Woosley \& Weaver(1995)]{WW95} Woosley, S.~E., \& Weaver, T.~A.\ 1995, ApJS, 101, 181 



\end{thebibliography}
\end{document}